\documentclass[11pt]{article}

\usepackage[dvips]{graphicx}
\usepackage[english]{babel}
\usepackage{psfrag}
\usepackage{slashed}

\usepackage{amsmath}

\newcommand{\slim}{\mskip 1.5mu}

\newcommand{\half}{ {\textstyle\frac{1}{2}} }

\newcommand{\mev}{\mbox{~MeV}}

\newcommand{\Tr}{\operatorname*{Tr}\nolimits}

\newcommand{\vev}[1]{\langle{#1}\rangle}

\newcommand{\widebar}[1]{\overline{#1}}
\newcommand{\Nvbar}{\overline{N}_{\!v}}

\newcommand{\Sym}[1]{\operatorname*{S}\displaylimits_{#1}}
\newcommand{\Asy}[1]{\operatorname*{A}\displaylimits_{#1}}
\newcommand{\subtr}[1]{\operatorname*{T}\displaylimits_{#1}}

\newcommand{\lrD}{{D^{\hspace{-0.8em}
      \raisebox{0.8ex}{$\scriptstyle\leftrightarrow$}}}{}}
\newcommand{\lD}{{D^{\hspace{-0.8em}
      \raisebox{0.8ex}{$\scriptstyle\leftarrow$}}}{}}
\newcommand{\rD}{{D^{\hspace{-0.8em}
      \raisebox{0.8ex}{$\scriptstyle\rightarrow$}}}{}}

\newcommand{\lpartial}{{\partial^{\hspace{-0.65em}
      \raisebox{0.8ex}{$\scriptstyle\leftarrow$}}}{}}
\newcommand{\rpartial}{{\partial^{\hspace{-0.65em}
      \raisebox{0.8ex}{$\scriptstyle\rightarrow$}}}{}}
\newcommand{\lrpartial}{{\partial^{\hspace{-0.65em}
      \raisebox{0.8ex}{$\scriptstyle\leftrightarrow$}}}{}}

\newcommand{\lrnab}{{\nabla^{\hspace{-0.8em}
      \raisebox{0.8ex}{$\scriptstyle\leftrightarrow$}}}{}}
\newcommand{\lnab}{{\nabla^{\hspace{-0.8em}
      \raisebox{0.8ex}{$\scriptstyle\leftarrow$}}}{}}
\newcommand{\rnab}{{\nabla^{\hspace{-0.8em}
      \raisebox{0.8ex}{$\scriptstyle\rightarrow$}}}{}}

\begin{document}
\begin{titlepage}
\begin{flushright}
\begin{tabular}{l}
DESY-06-196 \\
hep-ph/0611101
\end{tabular}
\end{flushright}

\vspace*{3em}

\begin{center}

\textbf{\LARGE Generalized parton distributions for the nucleon
  \\[0.3em] 
in chiral perturbation theory}

\vspace*{1.6cm}

{\large
M. Diehl$\slim{}^1$,
A. Manashov$\slim{}^{2,3}$
 and
A.~Sch\"afer$\slim{}^2$
}

\vspace*{0.4cm}

{\sl
$^1$ Deutsches Elektronen-Synchroton DESY, 22603 Hamburg, Germany \\
$^2$ Institut f\"ur Theoretische Physik, Universit\"at
  Regensburg, 93040 Regensburg, Germany \\
$^3$ Department of Theoretical Physics,  Sankt-Petersburg State
  University, St.-Petersburg, Russia
}

\vspace*{1.8cm}

\textbf{Abstract}\\[10pt]
\parbox[t]{0.9\textwidth}{
We complete the analysis of twist-two generalized parton distributions
of the nucleon in one-loop order of heavy-baryon chiral perturbation
theory.  Extending our previous study of the chiral-even isosinglet
sector, we give results for chiral-even isotriplet distributions and
for the chiral-odd sector.  We also calculate the one-loop corrections
for the chiral-odd generalized parton distributions of the pion.}

\vspace{1cm}

\end{center}

\end{titlepage}

\tableofcontents

\newpage

%%%%%%%%%%%%%%%%%%%%%%%%%%%%%%%%%%%
\section{Introduction}
%%%%%%%%%%%%%%%%%%%%%%%%%%%%%%%%%%%%

Generalized parton distributions (GPDs) provide a unified
parameterization of many different aspects of hadron physics
\cite{r1,r2,r3,r4}.  Understanding GPDs in detail is therefore
tantamount to understanding in large parts the internal structure of
hadrons. This motivates extensive experimental programs as well as
theoretical work. Details can be found in the reviews \cite{r5,r6,r7},
which emphasize the different types of physics encoded in these
quantities.  More recently it has been shown that interesting
information about the distribution of transversely polarized quarks in
a hadron is contained in GPDs associated with chiral-odd quark
operators \cite{Diehl:2005jf,Burkardt:2005hp}, for which there have
been relatively few studies so far.

The extraction of GPDs from experiment is a highly non-trivial task,
since in observables the distributions appear only within
convolutions.  These are relatively simple at leading order in the
strong coupling but become increasingly complex at higher orders, see
e.g.\ \cite{NNLO}.  In practice one therefore has to use
parameterizations of GPDs which are on one hand sufficiently flexible
to catch the physics and on the other hand contain only few
parameters.  In this context, the calculation of moments of GPDs in
lattice QCD is expected to become highly important in the future.

The lattice evaluation of these moments, parameterized by the form
factors of local matrix elements, is very similar to the case of the
usual electromagnetic form factors \cite{FF1}. The main limitation at
present is that lattice calculations with dynamical quarks can only be
done for unphysically heavy quarks and thus pions. The mass of the
pion affects however the spatial extent of the nucleon and hence its
form factors. Therefore, their extrapolation to the physical limit can
be fairly non-trivial, and simple linear extrapolations with respect
to $m_{\pi}^{}$ or $m_{\pi}^2$ could be quite inadequate. Progress in
this respect requires an analysis within chiral perturbation theory
(ChPT). We have presented such an analysis for the pion GPDs in
\cite{DMS1} and for nucleon GPDs in the chiral-even isosinglet sector
in \cite{DMS2}. In the present paper we extend this work to the
chiral-even isotriplet sector and the chiral-odd sector, giving
complete corrections at one-loop accuracy.  Calculations of a similar
scope have recently been reported in \cite{ando}, and we will compare
our results in detail.  There already exists a number of lattice
results for moments of GPDs, see \cite{Edwards:2006qx,Gockeler:2005vz}
and references therein. We do not include any ChPT fits to these in
the present paper, but leave them to future lattice studies.

Our paper is organized as follows. In Sections \ref{sect:even-gpds},
\ref{sect:chpt} and \ref{sect:tensor} we collect details about GPD
parameterizations, the operator product expansion, and heavy-baryon
ChPT that are needed in our analysis.  We proceed in each case by
constructing the operators within ChPT that match the relevant
twist-two operators in QCD, and by identifying the loop corrections
which contribute to a given form factor at relative order $O(q^2)$ in
the chiral expansion (Sections \ref{sect:isotriplet}, \ref{sect:ops}
and \ref{sect:tensor-ops}).  Results of the corresponding calculations
are given for the vector form factors in Section \ref{sect:vector},
for the axial form factors in Section \ref{sect:axial}, and for the
chiral-odd form factors in Section \ref{sect:tensor-results}. In
Section \ref{sect:gpd-results} all results are collected and rewritten
in terms of the usual parameterization of GPDs.  We summarize our main
findings in Section~\ref{sect:sum}.

%%%%%%%%%%%%%%%%%%%%%%%%%%%%%%%%%%%%%%%%%%%%%%%%%%%%%%%%%%%%%%%%%%%%%%%%%
\section{Chiral-even generalized parton distributions}
\label{sect:even-gpds}
%%%%%%%%%%%%%%%%%%%%%%%%%%%%%%%%%%%%%%%%%%%%%%%%%%%%%%%%%%%%%%%%%%%%%%%%%%

To begin with let us recall the definitions of generalized parton
distributions associated with chiral-even quark operators.  For the
distributions with definite isospin $I$ in a nucleon one can write
\begin{align}
  \label{quark-gpd}
& \int \frac{d \eta}{4\pi}\, e^{ix \eta(aP)}
\bigl\langle N_i(p') \bigl|\, 
  \bar{q}(-\half\eta a)\, \slashed{a} \slim 
  \tau^A q(\half\eta a) \,\bigr| N_j(p) \bigr\rangle
\nonumber \\
&\qquad = \tau^A_{ij}\, \frac{1}{2 aP}\, \bar{u}(p') \left[
  \slashed{a}\, H^{I}(x,\xi,t) +
  \frac{i \sigma^{\mu\nu} a_{\mu} \Delta_\nu}{2M} \, E^{I}(x,\xi,t)\, 
\right]  u(p) \,,
\displaybreak
\nonumber \\[2mm]
& \int \frac{d \eta}{4\pi}\, e^{ix \eta(aP)}
  \bigl\langle N_i(p') \bigl|\, 
  \bar{q}(-\half\eta a)\, \slashed{a}\gamma_5 \slim
  \tau^A q(\half\eta a) \,\bigr| N_j(p) \bigr\rangle
\nonumber \\
&\qquad = \tau^A_{ij}\, \frac{1}{2 aP}\, \bar{u}(p') \left[
  \slashed{a}\gamma_5\, \widetilde{H}^{I}(x,\xi,t) +
  \frac{a \Delta}{2M}\gamma_5\, \widetilde{E}^{I}(x,\xi,t) 
\right]  u(p) \,,
\end{align}
where $a$ is a light-like auxiliary vector, $M$ is the nucleon mass,
and we use the standard kinematical variables $P= \half (p+p')$,
$\Delta=p'-p$, $t=\Delta^2$ and $2\xi= -(\Delta a) /(P a)$.  Wilson
lines must be inserted between the quark fields if one is not working
in the light-cone gauge $(a A) =0$.  We combine the two-dimensional
unit matrix $\tau^0$ and the triplet of Pauli matrices $\vec{\tau}$ in
a four-vector $\tau^A = (\tau^0, \vec{\tau}\slim)$, with the matrices
acting on the isodoublet of quark fields $q$ or of nucleon states $N$.
The isosinglet distributions correspond to $A=0$ and the isotriplet
ones to $A=1,2,3$.  In terms of individual quark flavors in the proton
one has $H^{I=0}= H^u + H^d$ and $H^{I=1}= H^u - H^d$, with analogous
relations for the other distributions.

The Mellin moments of the GPDs in \eqref{quark-gpd} are related to the
matrix elements of the chiral-even local twist-two operators
\begin{align}
\label{operators}
\mathcal{O}^A_{\mu_1 \mu_2 \ldots \mu_n} &=
  \subtr{\mu_1 \ldots \mu_n} \Sym{\mu_1 \ldots \mu_n}
  \bar{q} \gamma_{\mu_1}
       i \lrD_{\mu_2} \ldots i \lrD_{\mu_n} \tau^A q \, , 
\nonumber \\
\widetilde{\mathcal{O}}^A_{\mu_1 \mu_2 \ldots \mu_n} &=
  \subtr{\mu_1 \ldots \mu_n} \Sym{\mu_1 \ldots \mu_n}
  \bar{q} \gamma_{\mu_1}\gamma_5\,
       i \lrD_{\mu_2} \ldots i \lrD_{\mu_n} \tau^A q
\end{align}
with $\lrD^\mu= \half (\rD^\mu-\lD^\mu)$.  Here $\subtr{}$ denotes the
subtraction of trace terms in the indicated Lorentz indices and
$\Sym{}$ denotes symmetrization, normalized as $\Sym{\mu_1\mu_2}
t^{\mu_1 \mu_2} = \half (t^{\mu_1 \mu_2} + t^{\mu_2 \mu_1})$.  Both
operations are conveniently implemented by contraction with the
auxiliary vector $a$,
\begin{align}
\label{defOa}
\mathcal{O}^A_n(a) &= a^{\mu_1}\ldots a^{\mu_n}\, 
  \mathcal{O}^A_{\mu_1\ldots\mu_n}\,, &
\widetilde{\mathcal{O}}^A_n(a) &= a^{\mu_1}\ldots a^{\mu_n}\,
  \widetilde{\mathcal{O}}^A_{\mu_1\ldots\mu_n} \,.
\end{align}
The local matrix elements can be parameterized as
\begin{align}
  \label{nucl-mat}
\vev{N_i(p')\slim
    |\, \mathcal{O}^A_n(a) \,|\slim N_j(p)}
&= \tau^A_{ij}\, \sum_{\substack{k=0\\{\mathrm{even}}}}^{n-1}
   (a P)^{n-k-1}\, (a  \Delta)^k\;
\bar u(p') \left[ \slashed{a}\, A_{n,k}^{I}(t)
     +\frac{i\sigma^{\mu\nu} a_{\mu}\Delta_\nu}{2M}\, B_{n,k}^{I}(t)
\right] u(p)
\nonumber \\
& \quad + \tau^A_{ij}\, \bmod(n+1,2)\, 
  (a\Delta)^n\, \frac{1}{M}\,\bar u(p') u(p)\, C_n^{I}(t)\,,
\nonumber\\[2mm]
\vev{N_i(p')\slim 
   |\, \widetilde{\mathcal{O}}^A_n(a) \,|\slim N_j(p)}
&= \tau^A_{ij}\, \sum_{\substack{k=0\\{\mathrm{even}}}}^{n-1}
   (a P)^{n-k-1}\, (a  \Delta)^k\;
\bar u(p') \left[ \slashed{a}\gamma_5\, \widetilde{A}_{n,k}^{I}(t)
           +\frac{a \Delta}{2M}\gamma_5\, \widetilde{B}_{n,k}^{I}(t)
\right] u(p) \,,
\end{align}
and the moments of the GPDs are given by
\begin{align}
  \label{gpd-mom}
\int_{-1}^1dx\, x^{n-1}\, H(x,\xi,t) &=
\sum_{\substack{k=0\\{\mathrm{even}}}}^{n-1}
(2\xi)^k\, A_{n,k}(t) + \mathrm{mod}(n+1,2)\,(2\xi)^{n} C_{n}(t)\,,
\nonumber \\
\int_{-1}^1dx\, x^{n-1}\, E(x,\xi,t) &=
\sum_{\substack{k=0\\{\mathrm{even}}}}^{n-1}
(2\xi)^k\, B_{n,k}(t) - \mathrm{mod}(n+1,2)\,(2\xi)^{n} C_{n}(t)\,,
\nonumber \\
\int_{-1}^1dx\, x^{n-1}\, \widetilde{H}(x,\xi,t) &=
\sum_{\substack{k=0\\{\mathrm{even}}}}^{n-1}
(2\xi)^k\, \widetilde{A}_{n,k}(t)\,,
\nonumber \\
\int_{-1}^1dx\, x^{n-1}\, \widetilde{E}(x,\xi,t) &=
\sum_{\substack{k=0\\{\mathrm{even}}}}^{n-1}
(2\xi)^k\, \widetilde{B}_{n,k}(t)\,,
\end{align}
where here and in the following we omit the isospin label $I$ when it
is not required.  The restriction to even $k$ in \eqref{nucl-mat} and
\eqref{gpd-mom} is a consequence of time reversal invariance.

To calculate the chiral corrections to the nucleon form factors in
heavy-baryon chiral perturbation theory we work in the Breit frame,
where $\boldsymbol{P}=0$.  The incoming and outgoing nucleons then
have opposite spatial momenta
$\boldsymbol{p}'=-\boldsymbol{p}=\boldsymbol{\Delta}/2$ and equal
energies, $p_0' =p_0^{\phantom{'}} =M\gamma$ with
\begin{equation}
  \gamma = \sqrt{1-\Delta^2/4M^2} \,.
\end{equation}
In terms of the velocity vector $v$, given by $v=(1,0,0,0)$ in the
Breit frame, the incoming and outgoing nucleon momenta are given by
$p=M\gamma v - \Delta/2$ and $p'=M\gamma v + \Delta/2$.  Note that $(v
\Delta)= (v S) =0$.  Dirac bilinears can be expressed in terms of the
velocity $v_\mu$ and the spin operator $S_\mu= \half\slim i
\sigma_{\mu\nu} \gamma_5\, v^\nu$.  Introducing the spinors
\begin{equation}
  \label{spinors}
u_v(p)=\mathcal{N}^{-1}\, \frac{1+\slashed{v}}{2}\,u(p) ,
\qquad
u_v(p')=\mathcal{N}^{-1}\, \frac{1+\slashed{v}}{2}\,u(p')
\end{equation}
with $\mathcal{N} = \sqrt{(1+\gamma)/2}$, the matrix elements in
\eqref{nucl-mat} can be rewritten as \cite{DMS2}
\begin{align}
  \label{OT}
\vev{N_i(p')\slim
  |\, \mathcal{O}^A_n(a) \,|\slim N_j(p)}
&= \tau^A_{ij}\, 
\sum_{k=-1}^{n-1} (M\gamma)^{n-k-2}\, (a v)^{n-k-1}\, (a \Delta)^{k}\, 
\nonumber \\
& \quad \times \bar u_v(p')\, \Big[ (a \Delta)\, E^{I}_{n,k+1}(t)
   +\gamma\,[(a S),(S\Delta)]\, M^{I}_{n,k}(t) \Big] \,u_v(p) \,,
\nonumber \\[2mm]
\vev{N_i(p')\slim
   |\, \widetilde{\mathcal{O}}^A_n(a) \,|\slim N_j(p)}
&= \tau^A_{ij}\, 
\sum_{k=0}^{n-1} (M\gamma)^{n-k-1}\, (a v)^{n-k-1}\, (a \Delta)^{k}\, 
\nonumber \\
& \quad \times \bar u_v(p') \left[
   2\gamma\slim (a S)\,{\widetilde E}^{I}_{n,k}(t)
   +\frac{(a \Delta) (S\Delta)}{2M^2}\,
    {\widetilde M}^{I}_{n,k}(t) \right] u_v(p) \,,
\end{align}
where due to time reversal invariance the terms with $E_{n,k+1}$ are
only nonzero for odd $k$, whereas those with $M_{n,k}$,
$\smash{\widetilde{E}}_{n,k}$ and $\smash{\widetilde{M}}_{n,k}$ are
only nonzero for even $k$.  The relation between the form factors in
\eqref{nucl-mat} and those in \eqref{OT} is
\begin{align}
  \label{ff-trafo}
E_{n,k}(t) &= A_{n,k}(t)+\frac{\Delta^2}{4M^2}B_{n,k}(t)
   \hspace{2em}\mbox{for~} k<n\,,&
E_{n,n}(t) &= {\gamma^2} \slim C_n(t)\,,
\nonumber \\[2mm]
M_{n,k}(t) &= A_{n,k}(t)+B_{n,k}(t)\,,
\nonumber \\[3mm]
\widetilde{E}_{n,k}(t) &= \widetilde{A}_{n,k}(t)\,,
\nonumber \\[2mm]
{\widetilde M}_{n,k}(t) &= (1+\gamma)^{-1}\widetilde{A}_{n,k}(t)
+\widetilde{B}_{n,k}\,,
\end{align}
which is readily inverted to
\begin{align}
  \label{ff-inverse}
A_{n,k}(t) &= \frac{1}{\gamma^2} \left[ E_{n,k}(t)
  - \frac{\Delta^2}{4M^2} M_{n,k}(t) \right] \,,
& B_{n,k}(t) &= \frac{1}{\gamma^2}\, \Big[ M_{n,k}(t)
  - E_{n,k}(t) \Big] \,,
\nonumber \\[2mm]
\widetilde{B}_{n,k}(t) &= \widetilde{M}_{n,k}(t)
  - (1+\gamma)^{-1} \widetilde{E}_{n,k}(t) \,.
\end{align}

%%%%%%%%%%%%%%%%%%%%%%%%%%%%%%%%%%%%%%%%%%%%%%%%%%%%%%%%%%%%%%%%%%%%%%%%
\section{Heavy-baryon ChPT}
\label{sect:chpt}
%%%%%%%%%%%%%%%%%%%%%%%%%%%%%%%%%%%%%%%%%%%%%%%%%%%%%%%%%%%%%%%%%%%%%%%%

To set our notation, let us briefly review the main ingredients of
chiral perturbation theory for heavy baryons, which is an effective
theory for the limit $q, m_\pi \ll M$, where $q$ is a generic
momentum.  To describe pions we use the nonlinear representation $U(x)
= \bigl[ u(x) \bigr]{}^2 = \exp\bigl[ i\pi^a(x)\slim \tau^a /F
\slim\bigr]$, where $F \approx 92 \mev$ is the pion decay constant in
the chiral limit.\footnote{%
Our convention is that uppercase indices of $\tau$ as in
\protect\eqref{quark-gpd} run from $0$ to $3$, whereas lowercase ones
run from $1$ to $3$.}
The explicit breaking of chiral symmetry by the quark masses is
implemented by the field $\chi(x)$.  We assume the isospin limit,
where one can replace $\chi(x) \to m^2\slim \tau^0$ with the bare pion
mass $m$.  We will not use external vector or axial vector fields
here.  The nucleon is described by the heavy-baryon field $N_v(x) =
\half (1+\slashed{v})\, e^{i M_0 (vx)} N(x)$, where $M_0$ is the bare
nucleon mass and $v$ the velocity vector.  The Fourier transform of
$N_v(x)$ depends on the residual nucleon momentum, given by the
original nucleon momentum minus $M_0 v$.  Important derived quantities
are the axial vector field
\begin{align}
  \label{u-def}
u_\mu &= i \bigl(u^\dagger \partial_\mu u-u\partial_\mu u^\dagger\bigr)
       = -\frac{1}{F}\,\partial_\mu \pi^a\tau^a +O(\pi^3) \,,
\\
\intertext{the connection}
  \label{Gamma-def}
\Gamma_\mu &= \frac{1}{2}
 \left(u^\dagger \partial_\mu u+u\partial_\mu u^\dagger\right)
  =\frac{i}{4F^2}\, \epsilon^{abc}\, \pi^a\, \partial_\mu\pi^b \tau^c
   +O(\pi^4) \,,
\end{align}
and
\begin{equation}
\chi_{\pm} = u^\dagger \chi u^\dagger \pm u \chi^\dagger u \,.
\end{equation}
Under global chiral transformations, described by unitary matrices
$V_L$ and $V_R$, the different fields transform as
\begin{align}
U &\to V_R^{\phantom{\dagger}}\slim U\slim V_L^\dagger \,, 
&
\chi &\to V_R^{\phantom{\dagger}}\slim \chi\slim V_L^\dagger \,,
\nonumber \\[0.1em]
u &\to V_R^{\phantom{\dagger}}\slim u\slim H^\dagger
     = H\slim u\slim V_L^\dagger \,,
&
N_v &\to H N_v \,,
\nonumber \\[0.1em]
\Gamma_\mu &\to  H\slim \Gamma_\mu H^\dagger 
           + H\slim \partial_\mu H^\dagger \,,
\end{align}
and $u_\mu$ and $\chi_\pm$ transform homogeneously as
\begin{align}
u_\mu     &\to H\slim u_\mu\slim     H^\dagger \,, &
\chi_\pm  &\to H\slim \chi_\pm\slim  H^\dagger \,.
\end{align}
The unitary matrix $H$ depends on $V_L$, $V_R$ and on $U(x)$ and
therefore has an $x$ dependence.  With the connection $\Gamma_\mu$ one
can construct the covariant derivative $\nabla_\mu$.  It acts as
$\nabla_\mu X = \partial_\mu X + \Gamma_\mu X$ on quantities like
$N_v$, which transform with a factor $H$ on their left, and as
$[\nabla_\mu, Y] = \partial_\mu Y + [\Gamma_\mu, Y]$ on quantities
like $u_\mu$, which transform with $H$ on the left and with
$H^\dagger$ on the right.  Corresponding derivatives acting to the
left are $Z\slim \lnab = Z\slim \lpartial - Z\slim \Gamma_\mu$ and
$[Y, \lnab] = Y\slim \lpartial - [Y, \Gamma_\mu]$, where $Z$
transforms with a factor $H^\dagger$ on its right.

The effective Lagrangian for the theory contains a pure pion piece and
a piece describing the nucleon and its interaction with pions,
$\mathcal{L}_{\mathrm{eff}}=\mathcal{L}_{\pi}+\mathcal{L}_{\pi N}$.
Expanding in powers of $q$ one has
\begin{align}
{\mathcal L}^{}_{\pi} &= {\mathcal L}_{\pi}^{(2)}
  +{\mathcal L}_{\pi}^{(4)} +\ldots\,, &
{\mathcal L}^{}_{\pi N} &= {\mathcal L}_{\pi N}^{(1)}
  +{\mathcal L}_{\pi N}^{(2)} +\ldots
\end{align}
with \cite{Gasser:1987rb,BFHM}
\begin{align}
  \label{Lpi}
\mathcal{L}^{(2)}_{\pi} &= \frac{F^2}{4}
  \Tr\bigl( u_\mu u^\mu + \chi_+ \bigr)\,,
\nonumber \\
\mathcal{L}^{(4)}_{\pi} &= 
    \frac{l_3}{16}\, \bigl( \Tr \chi_+ \bigr)^2
  + \frac{l_4}{16}\, \Big\{ 2 \Tr \chi_+ \Tr\slim (u_\mu u^\mu)
    + 2 \Tr\slim (\chi_-^2) 
    - \bigl( \Tr\chi_- \bigr)^2 \Big\} + \ldots \,,
\\
\intertext{and \protect\cite{BKM}}
  \label{LpiN}
\mathcal{L}^{(1)}_{\pi N} &=
\Nvbar\, \Big\{ i\slim (v \nabla) + g_0\slim (Su) \Big\} \,N_v\,,
\nonumber \\[0.15em]
\mathcal{L}^{(2)}_{\pi N} &= \Nvbar \left\{
  \frac{(v\nabla)^2 -\nabla^2}{2M_0}
  - \frac{ig_0}{2M_0}\, \bigl\{ (\nabla S), (v u)\bigr\}
  + c_1 \Tr \chi_+ \right.
\nonumber \\
& \hspace{2.2em} \left. {}
  + \Big( c_2 -\frac{g_0^2}{8M_0} \Big) (vu)^2 + c_3\, u_\mu u^\mu
  + \Big( c_4 + \frac{1}{4M_0} \Big) [S^\mu, S^\nu]\, u_\mu u_\nu
  \right\} N_v \,,
\end{align}
where $g_0$ is the nucleon axial-vector coupling in the chiral limit
and the $l_i$ and $c_i$ are further low-energy constants.  The terms
not displayed in $\mathcal{L}^{(4)}_{\pi}$ couple to at least four
pion fields and will not be needed in our calculations.

For calculating nucleon matrix elements in the Breit frame we need the
residual momenta of the incoming and outgoing nucleon,
\begin{align}
r  &= p  - M_0\slim v = wv - \Delta/2 \,, &
r' &= p' - M_0\slim v = wv + \Delta/2
\end{align}
with
\begin{equation}
  \label{w-def}
w = M (\gamma -1) + \delta M 
  = - \frac{\Delta^2}{8M} - 4c_1 m^2 + O(q^3) \,,
\end{equation}
where $\delta M = M-M_0$ is the nucleon mass shift.  Using the spinors
\eqref{spinors} one obtains a matrix element as
\cite{Steininger:1998ya}
\begin{equation}
  \label{matching}
\vev{p'|\mathcal{O}|p} = \mathcal{N}^2 Z_N\;
  \widebar{u}_v(p')\, G_{\mathcal{O}}(r',r)\, u_v(p) \,,
\end{equation}
where $G_{\mathcal{O}}(r',r)$ is the truncated Green function for
external heavy-baryon fields $\Nvbar$, $N_v$ and the operator
$\mathcal{O}$ in the effective theory.  $Z_N$ is the heavy-baryon
field renormalization constant,
\begin{equation}
  \label{ZN}
Z_N = 1 - \frac{3 m^2 g_0^2}{2\slim (4\pi F)^2}
  - \frac{9 m^2 g_0^2}{4\slim (4\pi F)^2} \log\frac{m^2}{\mu^2} 
  - 8m^2\slim d_{28}^{\slim r}(\mu) + O(q^3) \,,
\end{equation}
where $d_{28}^{\slim r}(\mu)$ is a low-energy constant in the
Lagrangian $\mathcal{L}_{\pi N}^{(3)}$ given in \cite{Fettes:1998ud}.

%%%%%%%%%%%%%%%%%%%%%%%%%%%%%%%%%%%%%%%%%%%%%%%%%%%%%%%%%%%%%%
\section{Chiral even isotriplet operators} 
\label{sect:isotriplet}
%%%%%%%%%%%%%%%%%%%%%%%%%%%%%%%%%%%%%%%%%%%%%%%%%%%%%%%%%%%%%%

%%%%%%%%%%%%%%%%%%%%%%%%%%%%%%%%%%%%%%%%%%%%%%%%
\subsection{Construction of effective operators}
\label{sect:operators}
%%%%%%%%%%%%%%%%%%%%%%%%%%%%%%%%%%%%%%%%%%%%%%%%

To find the operators in the effective theory which match the
quark-gluon operators \eqref{operators} in QCD we generalize the
construction of \cite{DMS2} to the isotriplet sector.  The relevant
effective operators contain a part ${\mathcal O}_\pi$ which involves
only pion fields (and couples to the nucleon via interactions from
$\mathcal{L}_{\pi N}$) and a part ${\mathcal O}_{\pi N}$ that is
bilinear in the nucleon field.  We thus have
\begin{align}
\label{OOa}
\mathcal{O}^A_n(a) &\,\cong\,
  \mathcal{O}^A_{n, \pi}(a) + \mathcal{O}_{n, \pi N}^A(a)\,, & 
\widetilde{\mathcal{O}}^A_n(a) &\,\cong\,
  \widetilde{\mathcal{O}}^A_{n, \pi}(a)
  + \widetilde{\mathcal{O}}_{n, \pi N}^A(a) \,,
\end{align}
where for the pure pion operators $\mathcal{O}^A_{n, \pi}(a)$ and
$\widetilde{\mathcal{O}}^A_{n, \pi}(a)$ we will use the form given
in~\cite{DMS1}.  The pion-nucleon operators $\mathcal{O}_{n, \pi
N}^A(a)$ and $\widetilde{\mathcal{O}}_{n, \pi N}^A(a)$ are
conveniently constructed by first matching the operators
\begin{align}
\label{isop}
\bigl( \mathcal{O}^{R}_{n}(a) \bigr)_{ij} &=
  \bar{q}_j\, \slashed{a}\, \frac{1+\gamma_5}{2}\, 
    (i a\lrD)^{n-1}\, q_i \,,
&
\bigl( \mathcal{O}^{L}_{n}(a) \bigr)_{ij} &=
  \bar{q}_j\, \slashed{a}\, \frac{1-\gamma_5}{2}\, 
    (i a\lrD)^{n-1}\, q_i \,,
\end{align}
where $i$ and $j$ are isospin indices.  They involve quarks of
definite chirality and transform as
\begin{align}
\label{isop-T}
\mathcal{O}^{R}_{n}(a) &\to
   V_R^{\phantom{\dagger}}\, \mathcal{O}^{R}_{n}(a)\slim V_R^\dagger
&
\mathcal{O}^{L}_{n}(a) &\to
   V_L^{\phantom{\dagger}}\slim \mathcal{O}^{L}_{n}(a)\slim V_L^\dagger
\end{align}
unter chiral rotations.  Parity transforms $\mathcal{O}^{R}_{n}(a)$
and $\mathcal{O}^{L}_{n}(a)$ into each other.  The corresponding
effective operators that are bilinear in the nucleon field can be
written in the form
\begin{align}
\bigl( Q^R_n(a) \bigr)_{ij}
 &= \bigl( \Nvbar\slim \mathcal{O}_1 u^\dagger \bigr)_{j}\,
    \bigl( u\slim \mathcal{O}_2 N_v \bigr)_{i} \,,
&
\bigl( Q^L_n(a) \bigr)_{ij}
 &= \bigl( \Nvbar\slim \mathcal{O}'_1 u \bigr)_{j}\,
    \bigl( u^\dagger\slim \mathcal{O}'_2 N_v \bigr)_{i} \,,
\end{align}
where $\mathcal{O}_{1}$, $\mathcal{O}_{2}$ transform like $u_\mu$
under chiral rotations and $\mathcal{O}'_{1}$, $\mathcal{O}'_{2}$ are
related to them by parity.  The vector and axial vector operators are
then readily obtained as
\begin{align}
\label{OLR}
\mathcal{O}^{A}_{n, \pi N}(a) 
&= \Tr \tau^A\slim \bigl\{ Q^{R}_{n}(a) + Q^{L}_{n}(a) \bigr\}
&
\widetilde{\mathcal{O}}^{A}_{n, \pi N}(a) 
&= \Tr \tau^A\slim \bigl\{ Q^{R}_{n}(a) - Q^{L}_{n}(a) \bigr\}
\end{align}
and will involve the combinations
\begin{equation}
  \label{tau-even-def}
\tau^A_{\,e \pm} = u^\dagger \tau^A u \pm u \tau^A u^\dagger \,,
\end{equation}
where the subscript $e$ indicates that they occur in chiral even
operators.  In the isosinglet case one has simply $\tau^0_{e+} =
2\tau^0$ and $\tau^0_{e-} = 0$, whereas the isotriplet combinations
\begin{align}
  \label{tau-even}
\tau^a_{e+} &= 2\tau^a + \frac{1}{F^2}\, \pi^b \bigl(
  \pi^a \tau^b - \pi^b \tau^a \bigr) + O(\pi^4)
&
\tau^a_{e-} &= - \frac{2}{F}\, \epsilon^{abc} \pi^b \tau^c
   + O(\pi^3)
\end{align}
involve an even or odd number of pion fields, respectively.  The
operators $\mathcal{O}_1$, $\mathcal{O}_2$ can be constructed from the
fields $u_\mu$ and $\chi_\pm$, and from the covariant derivatives
introduced in Section~\ref{sect:chpt}.  One can rearrange the
covariant derivatives in $Q^R_n(a)$ and $Q^L_n(a)$ to act either as
total derivatives $\partial_\mu$ on the product of all fields or in
the antisymmetric form $\lrnab_\mu = \half (\rnab_\mu - \lnab_\mu)$,
where $\rnab_\mu = \rpartial_\mu + \Gamma_\mu$ acts on the product of
all fields to the right and $\lnab_\mu = \lpartial_\mu - \Gamma_\mu$
on the product of all fields to the left.  The operators $Q^R_n(a)$
and $Q^L_n(a)$ are tensors having $n$ indices contracted with the
auxiliary vector $a$.  Other than $\partial_\mu$, $\lrnab_\mu$ and
$u_\mu$ these tensors can contain the vectors $v_\mu$ and $S_\mu$ and
the totally antisymmetric tensor.  The number of spin vectors can be
changed using the identities
\begin{align}
\label{SS}
\{S_\lambda,S_\mu\} &= \frac{1}{2} ( v_\lambda v_\mu-g_{\lambda\mu} ) \,, 
&
[ S_\lambda,S_\mu ] &= i\epsilon_{\lambda\mu\nu\rho}\, v^\nu S^\rho
&
S_\lambda &= -\frac{i}{2}\, 
  \epsilon_{\lambda\mu\nu\rho}\, v^\mu [ S^\nu, S^\rho ]
\end{align}
where our convention for the totally antisymmetric tensor is
$\epsilon_{0123}^{\phantom{0}}=1$.  For the operators under discussion
we chose a basis where $S_\mu$ appears at most linearly, or
quadratically as the commutator $[ S_\lambda,S_\mu ]$.  For counting
powers of $q$ one associates chiral dimension 1 to $\partial_\mu$,
$\lrnab_\mu$, $u_\mu$ and chiral dimension 2 to $\chi_{\pm}$.

We now make all factors of $(av)$ explicit and write
\begin{align}
\label{ON}
\mathcal{O}^A_{n, \pi N}(a)
= \sum_{k=0}^n M^{n-k-1}\, (a v)^{n-k}\, \mathcal{O}_{n,k}^A(a)\,,
\end{align}
where $\mathcal{O}_{n,k}^A(a)$ is free of factors $(a v)$.  For
contracting the $k$ vectors $a^\mu$ in $\mathcal{O}_{n,k}^A(a)$ one
can use $S_\mu$ only once, so that this operator contains at least
$k-1$ vectors $\partial_\mu$, $\smash{\lrnab_\mu}$ or $u_\mu$.  Thus
we can further decompose
\begin{equation}
  \label{ONK}
\mathcal{O}_{n,k}^A(a) =
\sum_{i=-1}^\infty M^{-i}\, \mathcal{O}_{n,k,i}^A(a) \,,
\end{equation}
where $\mathcal{O}_{n,k,i}^A(a)$ has chiral dimension $k+i$.  For
$\widetilde{\mathcal{O}}_{n, \pi N}^A(a)$ one has a decomposition in
full analogy to \eqref{ON} and \eqref{ONK}.

%%%%%%%%%%%%%%%%%%%%%%%%%%%%%%%%%%%%%%%%%%%%%%%%
\subsection{Power counting for tree and loop graphs}
\label{sect:power}
%%%%%%%%%%%%%%%%%%%%%%%%%%%%%%%%%%%%%%%%%%%%%%%%

As shown in \cite{DMS2} the chiral dimension of a graph with two
external nucleon legs and insertion of the operator
$\mathcal{O}_{n,k,i}^A(a)$ or $\widetilde{\mathcal{O}}_{n,k,i}^A(a)$
is
\begin{equation}
  \label{pow-count}
D_{k,i}= 2L +k+i
 +\sum_{j=1}^{N_{\pi}}\, \bigl(\dim{V_{\pi}}(j)-2 \bigr)
 +\sum_{j=1}^{N_{\pi N}}\, \bigl(\dim V_{\pi N}(j)-1 \bigr) \,,
\end{equation}
where $L$ is the number of loops (with $L=0$ for tree graphs).
$V_{\pi}(j)$ and $V_{\pi N}(j)$ respectively denote the $j$th vertex
from $\mathcal{L}_{\pi}$ and $\mathcal{L}_{\pi N}$ in the graph,
$N_{\pi}$ and $N_{\pi N}$ are the corresponding total numbers of
vertices, and $I_{\pi}$ and $I_N$ are the numbers of pion and nucleon
propagators.  Corrections to the nucleon propagator from higher orders
of $\mathcal{L}_{\pi N}$ are counted as a nucleon-nucleon vertex and
are accompanied by two (leading-order) nucleon propagators on either
side.  Notice that $r^\mu+r'^\mu = 2w v^\mu$ is of order $O(q^2)$ and
thus one order higher than the generic power associated with a
residual nucleon momentum.  A graph with chiral dimension $D_{k,i}$
can thus generate contributions to a nucleon matrix element of order
$O(q^{\slim d})$ with $d\ge D_{k,i}$.  Since
$\mathcal{O}_{n,k,i}^A(a)$ is accompanied by a factor $(av)^{n-k}$ in
\eqref{ON} it can only contribute to form factors with at least $n-k$
powers of $(av)$ in the decomposition \eqref{OT} of the nucleon matrix
element.  Taking into account the number $N_\Delta$ of factors
$(a\Delta)$ and $(S\Delta)$ in that decomposition, one can establish
the order in the chiral expansion to which a given operator can
contribute to a form factor.  The result is given in
Table~\ref{tab:chiral-order}.

Throughout this paper we refer to orders $O(q^d)$ in the chiral
expansion of a given \emph{form factor} rather than the expansion of
the corresponding \emph{matrix element}.  This is most convenient for
the problem at hand, since the chiral order of matrix elements
increases with the order $n$ of the operator, whereas the chiral order
of the form factors has as a natural point of reference the order
$O(q^0)$ from tree-level insertions of operators with the lowest
chiral dimension at given $n$.  

\begin{table}
\caption{\label{tab:chiral-order} Overview of contributions to the
  chiral even form factors.  The restriction in the second column is
  due to time reversal invariance.  $N_\Delta$ is the number of
  factors $(a\Delta)$ and $(S\Delta)$ in the decomposition
  \protect\eqref{OT}.  The indices of the operators must satisfy $l\ge
  k$ and $i\ge 0$, and the corresponding graphs contribute to the form
  factor at order $O(q^{\slim d})$ with $d \ge D_{l+1,i-1} - N_\Delta$
  and $D_{l+1,i-1}$ from \protect\eqref{pow-count}.}
\begin{center}
  \renewcommand{\arraystretch}{1.2}
  \begin{tabular}{cccc} \hline\hline
    form factor & $k$  & $N_\Delta$ & operators \\ \hline
    $E_{n,k+1}$ & odd  & $k+1$ & $\mathcal{O}_{n,l+1,i-1}$ \\
    $M_{n,k}$   & even & $k+1$ & $\mathcal{O}_{n,l+1,i-1}$ \\
    $\widetilde{E}_{n,k}$ & even & $k$ &
      $\widetilde{\mathcal{O}}_{n,l+1,i-1}$ \\
    $\widetilde{M}_{n,k}$ & even & $k+2$ &
      $\widetilde{\mathcal{O}}_{n,l+1,i-1}$ \\[0.2em]
    \hline\hline
  \end{tabular}
\end{center}
\end{table}

The contributions of the operators $\mathcal{O}_{n,k,i}^A(a)$ and
$\widetilde{\mathcal{O}}_{n,k,i}^A(a)$ at tree level are readily
evaluated.  The tree level graphs do not contain pions, so that one
can replace
\begin{align}
  \label{tree-rules}
u^\mu &\to 0, & \partial^\mu &\to i\Delta^\mu, & 
\lrnab^\mu &\to -i w v^\mu,
\nonumber \\
\tau^A_{\,e +} &\to 2\tau^A, & \tau^A_{\,e -} &\to 0, &
\chi_+ &\to 2 m^2 \tau^0, & \chi_- &\to 0.
\end{align}
Operators with $\lrnab_\mu$ do not contribute to the form factors at
leading order since $w$ is of order $O(q^2)$.  The different types of
higher-order contributions to the form factors are discussed in
Section~3.2 of \cite{DMS2}.  In the results we give for the form
factors, we lump them all into coefficients describing the $m^2$ and
$t$ corrections from tree graphs, except for the terms proportional to
$g_0^2$ in the expression \eqref{ZN} of the wave function
renormalization constant $Z_N$, which we combine with the terms due to
loop graphs.

\begin{figure}[t]
\psfrag{l}[c][c][0.8]{$l$}
\psfrag{a}[c][c][0.8]{$wv-\frac{\Delta}2$}
\psfrag{c}[c][c][0.8]{$wv+\frac{\Delta}2$}
\psfrag{e}[c][c][0.8]{$l+wv-\frac{\Delta}2$}
\psfrag{o}[c][c][0.8]{$l+wv+\frac{\Delta}2$}
\psfrag{aa}[b][b][1.0]{a} 
\psfrag{bb}[b][b][1.0]{b}
\psfrag{cc}[b][b][1.0]{c}
\centerline{\includegraphics[width=16cm]{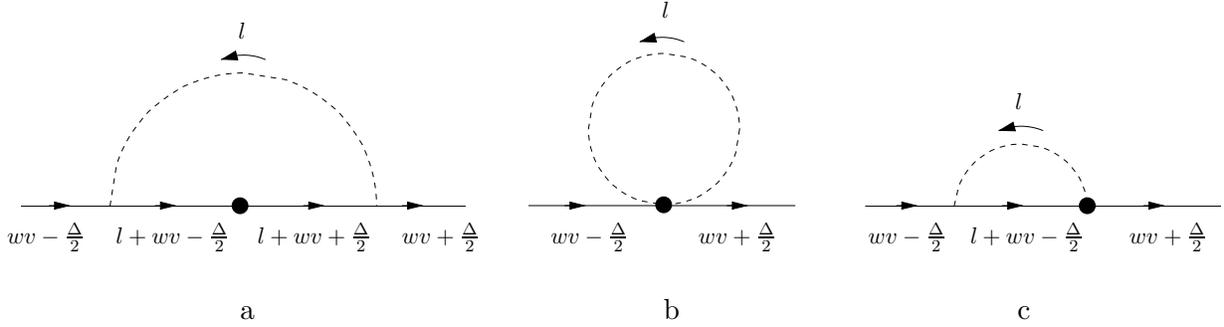}}
\caption{\label{fig-1} One-loop graphs with the insertion of a
pion-nucleon operator $\mathcal{O}_{n, \pi N}(a)$ or
$\widetilde{\mathcal{O}}_{n, \pi N}(a)$, which is denoted by a black
blob.  Not shown is the analog of graph c with residual momentum $l+wv
+\Delta/2$ of the intermediate nucleon line.}
\end{figure}

The one-loop graphs with pion-nucleon operator insertions are shown in
Fig.~\ref{fig-1}.  The construction of operators detailed in
Section~\ref{sect:operators} allows one to easily track the origin of
factors $\Delta_\mu$ which arise from a graph and must match the
factors in the form factor decomposition \eqref{OT}.  For this we use
that the denominators of the pion and nucleon propagators are $(l^2 -
m^2 +i\slim 0)$ and $(lv +w +i\slim 0)$, respectively, so that the
loop integration turns tensors $l_{\mu_1} \ldots l_{\mu_j}$ into
tensors constructed from $v_{\mu}$ and $g_{\lambda\mu}$.  We find that
with the leading-order (LO) interactions from $\smash{\mathcal{L}_{\pi
N}^{(1)}}$ and the next-to-leading (NLO) interactions from
$\smash{\mathcal{L}_{\pi N}^{(2)}}$ a factor $\Delta_\mu$ which is not
contracted to $\Delta^2$ (and hence can be contracted with $a^\mu$ or
$S^\mu$) can only originate from \cite{DMS2}
\begin{enumerate}
\item a total derivative $\partial_\mu$ in the operator insertion,
\item a term $(lv) (S\Delta)$ due to an NLO pion-nucleon vertex,
\item or a term $(l\Delta)$ due to an NLO nucleon propagator
  correction.
\end{enumerate}
We further find that two factors of $\Delta_\mu$ which are not
contracted to $\Delta^2$ can originate as $(l\Delta) (\Delta S)$ from
the NNLO pion-nucleon vertex generated by the term
\begin{equation}
  \label{LpiN3}
  -\frac{g_0}{4M_0^2}\, \Nvbar\, \Big\{ 
  (\lnab S) (u \rnab) + (\lnab u) (S \rnab) \Big\}\, N_v
\end{equation}
of the Lagrangian $\mathcal{L}_{\pi N}^{(3)}$ given in
\cite{Fettes:1998ud}.

In the next section we will see that $\mathcal{O}_{n,l+1,i-1}^A(a)$
and $\widetilde{\mathcal{O}}_{n,l+1,i-1}^A(a)$ with $i=0,1,2$ have at
most $l+i$ total derivatives $\partial_\mu$.  A one-loop graph with
insertion of such an operator and pion-nucleon interactions up to NNLO
must therefore satisfy
\begin{equation}
  \label{delta-cond}
l+i +\sum_{j=1}^{N_{\pi N}}\, \bigl(\dim V_{\pi N}(j)-1 \bigr)
\ge N_\Delta
\end{equation}
in order to produce the number $N_\Delta$ of factors $\Delta_\mu$
required to contribute to the form factors in
Table~\ref{tab:chiral-order}.  For $i>2$ and for pion-nucleon
interactions higher than NNLO this inequality is trivially fulfilled.
With the power counting established in the table, one then finds that
the one-loop contributions from pion-nucleon operators for all form
factors start at order $O(q^2)$.

%%%%%%%%%%%%%%%%%%%%%%%%%%%%%%%%%%%%%%%%%%%%
%%%%%%%%%%%%%%%%%%%%%%%%%%%%%%%%%%%%%%%%%%%%%

\begin{figure}[t]
\psfrag{a}[c][c][0.8]{$wv-\frac{\Delta}{2}$}
\psfrag{c}[c][c][0.8]{$wv+\frac{\Delta}{2}$}
\psfrag{e}[c][c][0.8]{$l+wv$}
\psfrag{b}[c][c][0.8]{$l+\frac{\Delta}{2}$}
\psfrag{d}[c][c][0.8]{$l-\frac{\Delta}{2}$}
\psfrag{f}[c][c][0.8]{$\Delta$}
\psfrag{aa}[b][b][1.0]{a} 
\psfrag{bb}[b][b][1.0]{b}
\psfrag{cc}[b][b][1.0]{c}
\centerline{\includegraphics[width=16cm]{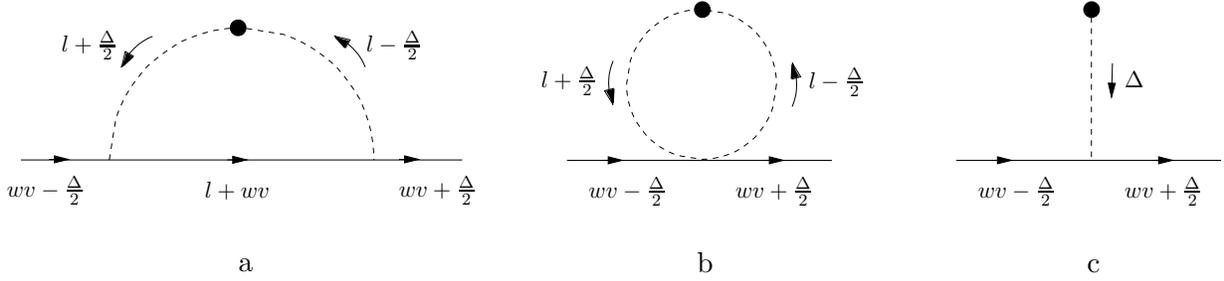}}
\caption{\label{fig-2} \textbf{a} and \textbf{b}: One-loop graphs with
the insertion of the pion operator $\mathcal{O}_{n, \pi}(a)$.
\textbf{c}: Tree graph with the insertion of the pion operator
$\widetilde{\mathcal{O}}_{n, \pi}(a)$.  The operator insertions are
denoted by black blobs.}
\end{figure}

Let us finally return to the contributions to the nucleon matrix
elements from the pure pion operators $\mathcal{O}^A_{n,\pi}(a)$ and
$\widetilde{\mathcal{O}}^A_{n,\pi}(a)$.  Their chiral dimension is
\cite{DMS2}
\begin{equation}
  \label{pion-dim}
D_\pi = 2L-1 + d_\pi
 +\sum_{j=1}^{N_{\pi}}\, \bigl(\dim{V_{\pi}(j)}-2 \bigr) 
 +\sum_{j=1}^{N_{\pi N}}\, \bigl(\dim V_{\pi N}(j)-1 \bigr)
\end{equation}
where $d_\pi \ge n$ is the chiral dimension of the pion operator.
Because of parity invariance the vector operators $\mathcal{O}_{n,
\pi}$ couple to 2 or more pions, whereas the axial vector operators
$\widetilde{\mathcal{O}}^A_{n,\pi}(a)$ couple to 1 but not 2 pions.
Starting at order $O(q^{n-k})$ the form factors $E_{n,k+1}$ and
$M_{n,k}$ thus receive corrections from the one-loop graphs shown in
Fig.~\ref{fig-2}a and b.  For isotriplet pion operators $n$ is odd due
to charge conjugation invariance.  Together with the time reversal
invariance constraints on the nucleon form factors, one thus finds
that the corrections to $E_{n,n-1}^{I=1}$ start at $O(q^2)$ and those
to $M_{n,n-1}^{I=1}$ at order $O(q)$, whereas for all other form
factors $E_{n,k+1}^{I=1}$ and $M_{n,k}^{I=1}$ they are at least of
order $O(q^3)$.

The axial vector operator $\widetilde{\mathcal{O}}^A_{n,\pi}(a)$
contributes to nucleon matrix elements starting with the tree level
graph in Fig~\ref{fig-2}c.  The $n$ vectors $a^\mu$ in the operator
are all contracted with derivatives acting on the pion field and hence
with $\Delta_\mu$ after evaluation of the graph.  The same is true for
the corresponding one-loop graphs.  One thus obtains only
contributions to the form factor $\widetilde{M}_{n,n-1}$, starting at
order $O(q^{-2})$ for the tree graph.  With two loops one has graphs
where three pions couple to the operator on one side and to the
nucleon line on the other.  Such graphs can contribute to other form
factors, but only starting at order $O(q^4)$.

%%%%%%%%%%%%%%%%%%%%%%%%%%%%%%%%%%%%%%%%%%%%%%%%%
%%%%%%%%%%%%%%%%%%%%%%%%%%%%%%%%%%%%%%%%%%%%%%%%%

%%%%%%%%%%%%%%%%%%%%%%%%%%%%%%%%%%%%%%%%%%%%%%%%%%%%%%%%%%%%%%
\section{Results for chiral-even isotriplet form factors} 
\label{sect:isotriplet-results}
%%%%%%%%%%%%%%%%%%%%%%%%%%%%%%%%%%%%%%%%%%%%%%%%%%%%%%%%%%%%%%

%%%%%%%%%%%%%%%%%%%%%%%%%%%%%%%%%%%%%%%%%%%%%%%%
\subsection{Relevant operators and graphs}
\label{sect:ops}
%%%%%%%%%%%%%%%%%%%%%%%%%%%%%%%%%%%%%%%%%%%%%%%%

With the method outlined in Section~\ref{sect:operators} one finds
vector operators
\begin{align}
  \label{Ovec}
\mathcal{O}^A_{n,k+1,-1} &=
  \widetilde{E}^{\slim I\slim (0)}_{n,k}\, (ia\partial)^{k}\,
    \Nvbar\, (aS)\, \tau^A_{\,e -} N_v + \ldots \,,
\nonumber \\[0.2em]
\mathcal{O}^A_{n,k+1,0} &=
  \half E^{\slim I\slim (0)}_{n,k+1}\, (ia\partial)^{k+1}\,
    \Nvbar\, \tau^A_{\,e +} N_v
- \half M^{\slim I\slim (0)}_{n,k}\, (ia\partial)^{k}\, i\partial_\mu
    \Nvbar\, [(aS), S^\mu] \tau^A_{\,e +} N_v + \ldots \,,
\nonumber \\[0.2em]
\mathcal{O}^A_{n,k+1,1} &= \tfrac{1}{4}
  \widetilde{M}^{\slim I\slim (0)}_{n,k}\, (ia\partial)^{k+1}\,
    i\partial_\mu \Nvbar\, S^\mu \tau^A_{\,e -} N_v + \ldots \,,
\end{align}
where the $\ldots$ stand for operators which have fewer total
derivatives and as in Section~\ref{sect:even-gpds} the isospin index
$I=0$ belongs to $A=0$ and $I=1$ to $A=1,2,3$.  The axial vector
operators are simply obtained by interchanging $\tau^A_{\,e +}$ and
$\tau^A_{\,e -}$,
\begin{align}
  \label{Oax}
\widetilde{\mathcal{O}}^A_{n,k+1,-1} &=
  \widetilde{E}^{\slim I\slim (0)}_{n,k}\, (ia\partial)^{k}\,
    \Nvbar\, (aS)\, \tau^A_{\,e +} N_v + \ldots \,,
\nonumber \\[0.2em]
\widetilde{\mathcal{O}}^A_{n,k+1,0} &=
  \half E^{\slim I\slim (0)}_{n,k+1}\, (ia\partial)^{k+1}\,
    \Nvbar\, \tau^A_{\,e -} N_v
- \half M^{\slim I\slim (0)}_{n,k}\, (ia\partial)^{k}\, i\partial_\mu
    \Nvbar\, [(aS), S^\mu] \tau^A_{\,e -} N_v + \ldots \,,
\nonumber \\[0.2em]
\widetilde{\mathcal{O}}^A_{n,k+1,1} &= \tfrac{1}{4}
  \widetilde{M}^{\slim I\slim (0)}_{n,k}\, (ia\partial)^{k+1}\,
    i\partial_\mu \Nvbar\, S^\mu \tau^A_{\,e +} N_v + \ldots \,.
\end{align}
Using the rules \eqref{tree-rules} and the decomposition \eqref{OT}
the coefficients in \eqref{Ovec} and \eqref{Oax} are easily identified
as the tree-level contributions to the form factors
$\widetilde{E}^I_{n,k}(t)$, $E_{n,k+1}^I(t)$, $M^I_{n,k}$ and
$\widetilde{M}^I_{n,k}$ at order $O(q^0)$.  Time reversal invariance
implies that ${E^{I\slim (0)}_{n,k+1}}$ is only nonzero for odd $k$,
whereas the other coefficients ${M^{I\slim (0)}_{n,k}}$,
${\widetilde{E}^{I\slim (0)}_{n,k}}$, ${\widetilde{M}^{I\slim
(0)}_{n,k}}$ are only nonzero for even $k$.  For $A=0$ we recover the
isosinglet operators constructed in~\cite{DMS2}.

According to \eqref{tau-even} insertions of an isotriplet operator with
$\tau^a_{e -}$ require at least one pion line in the graph and hence
do not contribute to nucleon matrix elements at tree level.  They
appear however in the one-loop graph shown in Fig.~\ref{fig-1}c.  When
the pion-nucleon vertex in this graph is taken at LO one finds zero,
because the loop integral is of the form
\begin{equation}
  \label{zero-loop}
\int d^{4-2\epsilon} l\;
  \frac{(S l)}{(lv +w +i\slim 0)\, (l^2-m^2+i\slim 0)} \,,
\end{equation}
whose numerator is proportional to $(S v) = 0$ after the integration.
Calculating the same graph to the next order, one finds that the
contribution from the NLO pion-nucleon vertex cancels the one with the
LO pion-nucleon vertex and an NLO nucleon propagator correction.  This
holds true for all operators with $\tau^a_{e -}$ in \eqref{Ovec} and
\eqref{Oax} and only requires that the operators does not introduce
any dependence on the loop momentum $l_\mu$ via $\lrnab_\mu$ or
$u_\mu$.  

The operators with $\tau^a_{e +}$ contribute at tree level and via the
loop graphs in Fig.~\ref{fig-1}a and b.  They are constructed such
that after the replacement $\partial_\mu \to i\Delta_\mu$ they match
the structure of the terms in the form factor decomposition
\eqref{OT}.  That structure can be changed in loop graphs only when
the spin vectors in the operator insertion are multiplied by spin
vectors from pion-nucleon vertices.  This is not the case for the
graph in Fig.~\ref{fig-1}b, which originates from the two-pion term in
the expansion \eqref{tau-even} of $\tau^a_{e+}$ and thus reproduces
the spin structure of the operator.  Let us show that it is not the
case either for the graph in Fig.~\ref{fig-1}a with LO pion-nucleon
vertices.  The numerator of the corresponding loop integral has the
form $(Sl)\slim \mathcal{O}\slim (Sl)$, where $\mathcal{O}$ contains
zero, one or two vectors $S_\mu$ and represents the spin structure of
the operator.  The loop integration turns a tensor $l_{\mu_1} \ldots
l_{\mu_j}$ into a combination of $v_{\mu}$ and $g_{\lambda\mu}$ and
thus $(Sl)\slim \mathcal{O}\slim (Sl)$ into $S^\rho\slim \mathcal{O}
S_\rho$.  This preserves the spin structure of $\mathcal{O}$ because
\begin{align}
S^\rho S_\rho &= \frac{1}{4} (1-d) \,, &
S^\rho\slim S^\alpha\slim S_\rho &= \frac{1}{4} (d-3)\, S^\alpha \,, &
S^\rho\slim [S^\alpha, S^\beta]\slim S_\rho = \frac{1}{4} (5-d)\,
[S^\alpha, S^\beta]
\end{align}
in $d$ dimensions.  We also need graphs with one LO and one NLO
pion-nucleon vertex, or with two LO vertices and an NLO nucleon
propagator correction.  Restricting ourselves to the terms producing
the required factors of $\Delta_\mu$ as discussed in
Section~\ref{sect:power}, we obtain numerators $(lv) (S\Delta)\,
\mathcal{O}\, (Sl)$ or $(l\Delta) (Sl)\, \mathcal{O}\, (Sl)$, which
give zero after loop integration.

In summary, the insertion of an operator from \eqref{Ovec} or
\eqref{Oax} into the graphs discussed so far either gives zero or
contributes only to the same form factor for which it already provides
the leading-order tree-level result.  This is however not true for the
graph in Fig~\ref{fig-1}a with one NNLO or two NLO pion-nucleon
interactions.  In this case one obtains terms $(S\Delta)\slim
\mathcal{O}\slim (S\Delta)$ after loop integration, which do change
the spin structure of $\mathcal{O}$.

%%%%%%%%%%%%%%%%%%%%%%%%%%%%%%%%%%%%%%%%%%%%%%%%
\subsection{Vector form factors}
\label{sect:vector}
%%%%%%%%%%%%%%%%%%%%%%%%%%%%%%%%%%%%%%%%%%%%%%%%

{}From Table~\ref{tab:chiral-order} it follows that $E_{n,k+1}^I$ and
$M^I_{n,k}$ can receive contributions from one-loop graphs with
insertion of operators $\mathcal{O}_{n,l+1,i-1}$ with $l\ge k$ and
$i\ge 0$.  With the additional condition \eqref{delta-cond} required
to produce enough factors of $\Delta_\mu$, we find that the form
factors receive corrections of order $O(q^2)$ from graphs with LO
pion-nucleon vertices and insertion of the operator
$\mathcal{O}_{n,k+1,0}$, which already gives the tree-level
contributions at order $O(q^0)$.  By power counting one could also
have order $O(q^2)$ contributions from graphs with insertion of
$\mathcal{O}_{n,k+2,-1}$ or $\mathcal{O}_{n,k+1,-1}$ and pion-nucleon
interactions at LO or NLO, respectively, but these vanish because the
relevant operators come with $\tau^A_{\,e -}$.  The one-loop
corrections from pion-nucleon operators to $E_{n,k}^{I=1}(t)$ and
$M_{n,k}^{I=1}(t)$ are then found to be
\begin{align}
  \label{ME-nucl}
E_{n,k}^{I=1\slim (0)}
\left(1-\frac{m^2}{(4\pi F)^2}
\left[(3g_A^2+1)\log\frac{m^2}{\mu^2}+2g_A^2\right]\right)
+ O(q^3) \,,
\nonumber \\[0.3em]
M_{n,k}^{I=1\slim (0)}
\left(1-\frac{m^2}{(4\pi F)^2}
\left[(2g_A^2+1)\log\frac{m^2}{\mu^2}+2g_A^2\right]\right) 
+ O(q^3) \,,
\end{align}
respectively, where we have replaced the bare axial coupling $g_0$ by
its physical value $g_A$ as is permissible within the precision of our
result.  Likewise, we could replace the bare pion decay constant $F$
and bare pion mass $m$ by their physical values $F_\pi$ and $m_\pi$
(we refrain from doing so for ease of notation).
The contributions with $g_A^2$ in \eqref{ME-nucl} are due to the graph
in Fig.~\ref{fig-1}a and the nucleon wave function renormalization,
and the contributions without $g_A^2$ come from the tadpole graph in
Fig.~\ref{fig-1}b.  As in \cite{DMS2} we use the renormalization
scheme of \cite{GL}, subtracting $1/\epsilon+\log(4\pi)+\psi(2)$ for
each $1/\epsilon$ pole in $4-2\epsilon$ dimensions.

The form factors $E_{n,k+1}^I(t)$ and $M^I_{n,k}$ also receive chiral
corrections from loop graphs with pion operator insertions.  In the
notation of \cite{DMS1} the isotriplet operators with lowest chiral
dimension are\footnote{%
The normalization of the twist-two operators \protect\eqref{operators}
used here agrees with the one in \protect\cite{DMS2} and differs from
that in \protect\cite{DMS1} by a factor of 2.  The coefficients
$\tilde{b}_{n,k}$ have the same normalization here and in
\protect\cite{DMS1}.}
\begin{align}
\label{sv-n}
\mathcal{O}_{n, \pi}^{a}(a) &= 
  2\tilde b_{n,n-1}\, (i a\partial)^{n-1} (aV^a)
\nonumber \\
 &\quad + 2 iF^2\epsilon^{abc}\,
\sum_{\substack{k=0\\ \mathrm{even}}}^{n-3}
   \tilde{b}_{n,k}\, (i a\partial)^k \left[
(aL^{b})\,(2i a\lrpartial)^{n-k-2} (aL^{c})+
(aR^{b})\,(2i a\lrpartial)^{n-k-2} (aR^{c}) \right] ,
\end{align}
where
\begin{align}
V^a_\mu &= -\half {i}\slim F^2 \bigl( L^a_\mu + R^a_\mu \bigr) \,,
&
L^a_\mu\slim \tau^a &= U^\dagger\slim \partial_\mu U  \,,
&
R^a_\mu\slim \tau^a &= U\slim \partial_\mu U^\dagger \,.
\end{align}
To extract the terms coupling to two pions we use the expansion
$L^a_\mu = i\partial_\mu \pi^a /F + i\epsilon^{abc} \pi^b \partial_\mu
\pi^c /F^2 + O(\pi^3)$ and its analog for $R^a_\mu$, obtained by
changing the sign of the pion field,\footnote{%
We note that the sign of the term with $\epsilon^{abc}$ in eq.~(32) of
\protect\cite{DMS1} is incorrect.}
and obtain
\begin{align}
\label{sv-nexp}
\mathcal{O}_{n, \pi}^{a}(a) &=
 -2i\epsilon^{abc}\, \Bigg\{ \tilde b_{n,n-1}\,
(i a\partial)^{n-1} \pi^b (ia\partial\slim \pi^c) -
2 \sum_{\substack{k=0\\ \mathrm{even}}}^{n-3}
  \tilde{b}_{n,k}\, (i a\partial)^k \Big[ (i a\partial\slim \pi^b)\, 
  (2i a\lrpartial)^{n-k-2}\, (i a\partial\slim \pi^c) \Big] \Bigg\}
\nonumber \\
 &\quad + O(\pi^4) \,.
\end{align}
Using the relations
\begin{align}
4 (i a\partial\slim \pi^b)\, (i a\partial\slim \pi^c)
 &= (ia\partial)^2 \pi^{b}\pi^c - \pi^{b}\, (2i a\lrpartial)^2\, \pi^c 
\,, &
2 \epsilon^{abc} \pi^b\, (ia\partial\slim \pi^c)
 &= \epsilon^{abc} \pi^b\, (2ia\lrpartial)\slim \pi^c
\end{align}
we can rewrite this as
\begin{align}
\label{qa}
\mathcal{O}_{n, \pi}^{a}(a) &= -i\epsilon^{abc}\,
\sum_{\substack{k=0\\ \mathrm{even}}}^{n-1}
A_{n,k}^{\pi\slim (0)}\, (i a\partial)^k
\Big[ \pi^b\,(i a\lrpartial)^{n-k}\, \pi^c \Big] + O(\pi^4) \,,
\end{align}
where $A_{n,k}^{\pi\slim (0)} = 2^{n-k} \bigl( \tilde b_{n,k}-\tilde
b_{n,k-2} \bigr)$ with $\tilde{b}_{n,-2}=0$.  The coefficients
$A_{n,k}^{\pi\slim (0)}$ represent the chiral limit of the form
factors $A_{n,k}^\pi(t)$ which parameterize the moments of the pion
GPD as \cite{DMS1}
\begin{align}
  \label{pion-gpds}
\int_{-1}^1 dx\, x^{n-1} H_\pi^{I}(x,\xi,t)
=\sum_{\substack{k=0\\ \mathrm{even}}}^n (2\xi)^k A_{n,k}^\pi(t) \,,
\end{align}
where in terms of quark flavors in a $\pi^+$ one has $H_\pi^{I=0} =
H_\pi^{u} + H_\pi^{d}$ and $H_\pi^{I=1} = H_\pi^{u} - H_\pi^{d}$.
Because of isospin and charge conjugation symmetry one has $I=1$ for
odd $n$ and $I=0$ for even $n$ and therefore can omit the isospin
index $I$ in $A_{n,k}^\pi$.

As discussed after \eqref{pion-dim}, the graphs in Fig.~\ref{fig-2}a
and b with insertion of $\mathcal{O}_{n, \pi}^{a}(a)$ give rise to
corrections which start at order $O(q^2)$ for $E^{I=1}_{n,n-1}$ and at
order $O(q)$ for $M^{I=1}_{n,n-1}$.  Together with \eqref{ME-nucl} and
with terms due to tree level operator insertions, the complete results
to order $O(q^2)$ read
\begin{align}
\label{EM-full}
E_{n,k}^{I=1}(t) &=
E_{n,k}^{I=1\slim (0)}
\left(1-\frac{m^2}{(4\pi F)^2}
\left[(3g_A^2+1) \log\frac{m^2}{\mu^2} + 2g_A^2\right]\right)
\nonumber \\[0.2em]
 &\quad + \delta_{k,n-1}^{\phantom{0}}\, 
          E_{n,n-1}^{I=1\slim (2,\pi)}(t)
+ E_{n,k}^{I=1\slim (2,m)}\slim m^2 + E_{n,k}^{I=1\slim (2,t)}\slim t
+ O(q^3) \,,
\nonumber \\[0.3em]
M_{n,k}^{I=1}(t) &=
M_{n,k}^{I=1\slim (0)}
\left(1-\frac{m^2}{(4\pi F)^2}
\left[(2g_A^2+1) \log\frac{m^2}{\mu^2} + 2g_A^2\right]\right)
\nonumber \\[0.2em]
 &\quad + \delta_{k,n-1}^{\phantom{0}}\,
    \Big[ M_{n,n-1}^{I=1\slim (1,\pi)}(t)
        + M_{n,n-1}^{I=1\slim (2,\pi)}(t) \Big]
+ M_{n,k}^{I=1\slim (2,m)}\slim m^2 + M_{n,k}^{I=1\slim (2,t)}\slim t
+ O(q^3) \,,
\end{align}
where the contributions
\begin{align}
  \label{EM-pi}
E_{n,n-1}^{I=1\slim (2,\pi)}(t) &= \frac{1}{2(4\pi F)^2}
  \sum_{\substack{j=1\\ \mathrm{odd}}}^n 2^{-j} j\,
  A_{n,n-j}^{\pi\slim (0)} \, \Bigg\{
  4 g_A^2\slim m^2\log\frac{m^2}{\mu^2}
\nonumber \\
 &\quad + \int_{-1}^1 d\eta\, \eta^{j-1} \left[
  g_A^2\, (2m^2 - t) \left( \log\frac{m^2(\eta)}{\mu^2}+1 \right)
  - (g_A^2-1)\, m^2(\eta)\log\frac{m^2(\eta)}{\mu^2} \right] \Bigg\} \,,
\nonumber \\[0.4em]
M_{n,n-1}^{I=1\slim (1,\pi)}(t) &= 
 - \frac{2\pi\slim M g_A^2}{(4\pi F)^2}
\sum_{\substack{j=1\\ \mathrm{odd}}}^n 2^{-j} j\,
  A_{n,n-j}^{\pi\slim (0)} \int_{-1}^1 d\eta\, \eta^{j-1}\, m(\eta)
\end{align}
with
\begin{equation}
m^2(\eta)=m^2- \frac{t}{4}\slim (1-\eta^2)
\end{equation}
are due to graphs with pion operator insertions and LO pion-nucleon
vertices.  The order $O(q^2)$ correction
\begin{align}
  \label{M2-pi}
& M_{n,n-1}^{I=1\slim (2,\pi)}(t) = - \frac{1}{2(4\pi F)^2}
  \sum_{\substack{j=1\\ \mathrm{odd}}}^n 2^{-j} j\,
  A_{n,n-j}^{\pi\slim (0)} \, \Bigg\{
  4 g_A^2\slim m^2\log\frac{m^2}{\mu^2}
\nonumber \\
& \quad + \int_{-1}^1 d\eta\, \eta^{j-1} \left[
  g_A^2\, \bigl( 2m^2 - t \bigr) 
          \left( \log\frac{m^2(\eta)}{\mu^2}+1 \right)
  + \bigl( g_A^2 -1-4M c_4 \bigr)\,
    m^2(\eta) \log\frac{m^2(\eta)}{\mu^2} \right] 
\Bigg\}
\end{align}
is due to graphs with one NLO pion-nucleon vertex or nucleon
propagator correction, as well as graphs with LO vertices and the
subleading part $w v^\mu$ of the residual nucleon momenta, cf.\ the
discussion after \eqref{pow-count}.  The terms proportional to $g_A^2$
in \eqref{EM-pi} and \eqref{M2-pi} are due to the graph in
Fig.~\ref{fig-2}a, and the other terms to the graph in
Fig.~\ref{fig-2}b.  
Our expressions \eqref{EM-full} and \eqref{EM-pi} agree with the
results in \cite{ando}, where the order $O(q^2)$ corrections to
$E_{n,n-1}^{I=1}$ and the order $O(q)$ corrections to $M_{n,n-1}^{I=1}$
are given.

%%%%%%%%%%%%%%%%%%%%%%%%%%%%%%%%%%%%%%%%%%%%%%%%
\subsection{Axial form factors}
\label{sect:axial}
%%%%%%%%%%%%%%%%%%%%%%%%%%%%%%%%%%%%%%%%%%%%%%%%

Using Table~\ref{tab:chiral-order} and the condition
\eqref{delta-cond}, one readily finds that the chiral corrections of
order $O(q^2)$ to the form factor $\tilde{E}_{n,k}^I$ are obtained
from graphs with LO vertices and insertion of the operator
$\widetilde{\mathcal{O}}_{n,k+1,-1}$, which already gives the
tree-level contributions at order $O(q^0)$.  Together with
higher-order tree level insertions we get
\begin{align}
\label{Ea-cor}
\widetilde{E}_{n,k}^{I=1}(t)
 &= \widetilde{E}_{n,k}^{I=1\slim (0)}
\left(1-\frac{m^2}{(4\pi F)^2}
\left[(2g_A^2+1) \log\frac{m^2}{\mu^2} + {g_A^2}\right]\right)
\nonumber \\[0.3em]
&\quad 
+ \widetilde{E}_{n,k}^{I=1\slim (2,m)}\slim m^2 
+ \widetilde{E}_{n,k}^{I=1\slim (2,t)}\slim t + O(q^3) \,,
\end{align}
in agreement with \cite{ando}. 
The discussion of contributions to $\widetilde{M}_{n,k}^I$ is more
involved; for the isotriplet case it proceeds in close analogy to the
isosinglet case analyzed in \cite{DMS2}.  According to
Table~\ref{tab:chiral-order} and the condition \eqref{delta-cond}, one
obtains order $O(q^2)$ corrections from graphs with insertion of
$\widetilde{\mathcal{O}}_{n,k+1,1}$ and LO vertices.  Further
corrections are due to graphs with insertion of
$\widetilde{\mathcal{O}}_{n,k+1,-1}$ and two NLO pion-nucleon vertices
or nucleon propagator corrections, or with one NNLO pion-nucleon
vertex generated by \eqref{LpiN3}.  Graphs with the same operator
insertion and two loops or one loop and a pion propagator correction
from $\mathcal{L}_{\pi}^{(4)}$ could contribute by power counting but
do not produce the required factors of $\Delta_\mu$ (see \cite{DMS2}).
Graphs with insertion of $\widetilde{\mathcal{O}}_{n,k+2,0}$ or
$\widetilde{\mathcal{O}}_{n,k+1,0}$ and pion-nucleon interactions at
LO or NLO give zero because these operators involve $\tau^a_{e -}$, as
discussed after \eqref{zero-loop}.  Graphs involving
$\widetilde{\mathcal{O}}_{n,k+2,-1}$ and NLO pion-nucleon interactions
do not contribute to $\widetilde{M}_{n,k}^I$ due to time reversal
invariance, since the operator is only nonzero for odd $k$ and the
form factor only for even $k$.  Finally, graphs with insertion of
$\widetilde{\mathcal{O}}_{n,k+3,-1}$ and LO pion-nucleon vertices
contribute to $\widetilde{E}_{n,k+2}^I$ but not to
$\widetilde{M}_{n,k}^I$ as discussed at the end of
Section~\ref{sect:ops}.

Together with higher-order tree-level insertions, the one-loop graphs
with $\widetilde{\mathcal{O}}_{n,k+1,1}$ or
$\widetilde{\mathcal{O}}_{n,k+1,-1}$ thus give the full result at
order $O(q^2)$ for the form factors with $k < n-1$,
\begin{align}
\label{Ma-cor}
\widetilde{M}_{n,k}^{I=1}(t)
 &= \widetilde{M}_{n,k}^{I=1\slim (0)}
\left(1-\frac{m^2}{(4\pi F)^2}
\left[(2g_A^2+1) \log\frac{m^2}{\mu^2} + {g_A^2}\right]\right)
\nonumber \\[0.3em]
&\quad 
+ \widetilde{E}_{n,k}^{I=1\slim (0)}\, \frac{m^2 g_A^2}{3 (4\pi F)^2}
  \log\frac{m^2}{\mu^2}
+ \widetilde{M}_{n,k}^{I=1\slim (2,m)}\slim m^2 
+ \widetilde{M}_{n,k}^{I=1\slim (2,t)}\slim t + O(q^3) \,.
\end{align}

The form factors $\smash{\widetilde M_{n,n-1}^{I=1}}$ require a
separate discussion because they receive a contribution starting at
order $O(q^{-2})$ from the one-pion exchange graph in
Fig.~\ref{fig-2}c, as discussed at the end of
Section~\ref{sect:power}.  The relevant operator is given by
\begin{equation}
  \label{OTnpi}
\widetilde{\mathcal{O}}_{n, \pi}^{a}(a) = 
  2\tilde b_{n,n-1}\, (i a\partial)^{n-1} (aA^a) \,
\left[\, 1 + \left( \frac{l_4^r}{2 F^2} + \tilde{c}_n^{} \right) 
             \Tr \chi_+ \,\right]
+ O\bigl(q^{n+4}\bigr)
\end{equation}
with odd $n$ and
\begin{equation}
A^a_\mu = -\half {i}\slim F^2 \bigl( R^a_\mu - L^a_\mu \bigr) \,,
\end{equation}
where $\tilde{b}_{n,n-1}$ is the same as in \eqref{sv-n} because of
parity invariance.  $l_4^r$ is the renormalized low-energy constant
from the pion Lagrangian \eqref{Lpi} and appears in the expression of
the axial current,
\begin{equation}
  \label{ax-curr}
\frac{1}{2}\slim \bar{q} \gamma_\mu \gamma_5 \tau^a q  
\,\cong\, A_\mu^a\,
   \left[\, 1 + \frac{l_4^r}{2 F^{2}} \Tr\chi_+ \,\right] + O(q^5) \,,
\end{equation}
so that $\tilde{b}_{1,0} = 1$ and $\tilde{c}_1 = 0$.  One can readily
derive \eqref{ax-curr} by coupling the Lagrangian to an external
isovector axial field $a_\mu$ as usual \cite{BKM}, which implies
$u_\mu = i \bigl(u^\dagger \partial_\mu u - u \partial u^\dagger
\bigr) + u^\dagger a_\mu u + u\slim a_\mu u^\dagger$.  As an aside we
note that the correction with $l_4^r$ in the axial current
\eqref{ax-curr} would be different if one used the pion Lagrangian
from \cite{GL}, where the term involving this low-energy constant
reads
\begin{equation}
  \label{GL-Lpi}
\frac{l_4}{4} \Tr\slim
  \Bigl[ \slim (\partial_\mu \chi^\dagger) (\partial^\mu U)
             + (\partial_\mu \chi) (\partial^\mu U^\dagger) \slim\Bigr]
= \frac{l_4}{8}\,
  \Bigl\{ \Tr\chi_+ \Tr\slim (u_\mu u^\mu)
        + 2i \Tr\slim \bigl[ \nabla_\mu, \chi_- \bigr]\slim
          u^\mu \slim\Bigr\} \,.
\end{equation}
In the present work we follow Ref.~\cite{BFHM} and use the Lagrangian
\eqref{Lpi} from \cite{Gasser:1987rb}.  It differs from \eqref{GL-Lpi}
by total derivative terms and terms that vanish by the equation of
motion.  With the full Lagrangian given by $\mathcal{L}_{\pi} +
\mathcal{L}_{\pi N}$ the equation of motion for the pion field
involves terms bilinear in the nucleon field,\footnote{%
With an arbitrary matrix $X$ in isospin space and $\widetilde{X} = X -
\half \Tr X$, the leading-order equation of motion reads
\begin{equation*}
 2i \Tr \bigl[ \nabla_\mu, u^\mu \bigr] X
 + \Tr \chi_- \slim \widetilde{X}
 = F^{-2}\; \Nvbar\slim \bigl[ (vu) + 4ig_0\slim (S\nabla),
            \widetilde{X} \slim\bigr] N_v
  - 4i g_0 F^{-2}\, \partial^\mu\slim 
    \bigl(\slim \Nvbar\slim S_\mu \widetilde{X} N_v \bigr) \,.
\end{equation*}
} 
so that a change of $\mathcal{L}_{\pi}$ using the equation of motion
induces a corresponding change in $\mathcal{L}_{\pi N}$.  We also
refer to the discussion in \cite{Ecker:1995rk}.

The coefficients $\tilde{b}_{n,n-1}$ and $\tilde{c}_{n}$ in
\eqref{OTnpi} appear in the moments of the twist-two pion distribution
amplitude $\phi_\pi(x)$, which are defined by
\begin{equation}
\int \frac{d \eta}{2\pi}\, e^{ix \eta(ap)}
\bigl\langle \pi^b(p) \bigl|\, 
  \bar{q}(-\eta a)\, \slashed{a} \gamma_5 \slim 
  \tau^a q(\eta a) \,\bigr| 0 \bigr\rangle 
= - i \delta^{ab} F_\pi \phi_\pi(x)
\end{equation}
and 
\begin{equation}
  \label{piDAmom}
B_n^\pi = 2^{-n} \int_{-1}^1 dx\, x^{n-1} \phi_\pi(x) \,,
\end{equation}
so that
\begin{equation}
\langle \pi^b(p) |\, \widetilde{\mathcal{O}}_{n, \pi}^{a}(a) 
  | 0\rangle = -2i \delta^{ab} (ap)^n F_\pi^{}\slim B_n^\pi \,.
\end{equation}
Calculating the leading chiral correction to this matrix element one
finds that $l_4^r$ only appears in the expression of the pion decay
constant,
\begin{equation}
  \label{Fpi}
F_\pi = F \left( 1 - \frac{m^2}{(4\pi F)^2} \log\frac{m^2}{\mu^2}
  + \frac{m^2}{F^2}\, l_4^r(\mu) + O(m^4) \right) \,,
\end{equation}
and $\tilde{c}_n$ only in the correction to the moments
\begin{equation}
  \label{Bn}
B_n^\pi = \tilde{b}_{n,n-1}^{}\, 
  \bigl( 1 + 4 m^2 \tilde{c}_n^{} \bigr) + O(m^4) \,.
\end{equation}
The definition of the pion decay constant implies $B_1^\pi = 1$ to all
orders in the chiral expansion.

Returning to the nucleon form factors $\widetilde{M}_{n,n-1}^{I=1}$,
one readily finds their lowest-order contribution to be
\begin{equation}
  \label{pion-tree}
\widetilde{M}_{n,n-1}^{I=1\slim (-2)}(t)
  = B_{n\phantom{j}}^{\pi\slim (0)}\, \frac{4 M^2 g_0}{m^2 - t}
\end{equation}
with $B_n^{\pi (0)} = \tilde{b}_{n,n-1}$.  The leading corrections to
this come from a number of higher-order operator insertions and loop
graphs.  A tadpole insertion in the pion line of the graph in
Fig.~\ref{fig-2}c and the propagator correction from the $l_3$ term in
the pion Lagrangian \eqref{Lpi} result in a shift $m^2\to m_\pi^2$ in
\eqref{pion-tree}, where
\begin{equation}
m_\pi^2 = m^2 \left( 1
  + \frac{m^2}{2 (4\pi F)^2} \log\frac{m^2}{\mu^2}
  + \frac{2 m^2}{F^2}\, l_3^r(\mu) + O(m^4) \right) \,.
\end{equation}
The pion propagator correction from the $l_4$ term in \eqref{Lpi}
cancels against the $l_4$ term from the operator \eqref{OTnpi}.  A
tadpole insertion at the operator vertex gives a chiral logarithm as
in \eqref{Fpi}.  Further corrections are due to loop corrections to
the pion-nucleon vertex, to tree-level insertions from
$\mathcal{L}_{\pi N}^{(3)}$, and to the factors $\mathcal{N}^2$ and
$Z_N$ in the matching formula \eqref{matching}.  Together with the
tree-level insertion of the pion-nucleon operator
$\mathcal{\widetilde{O}}_{n,n,1}^a$ from \eqref{Oax} we obtain
\begin{align}
  \label{Mnn-1}
\widetilde M_{n,n-1}^{I=1}(t) =
B_{n}^{\pi\slim (0)}\, \frac{4 M^2 g_0}{m_\pi^2 - t} \,
& \left( 1 - \frac{m^2}{(4\pi F)^2}
  \left[(2g_0^2+1) \log\frac{m^2}{\mu^2} + {g_0^2} \slim\right] \right.
\nonumber \\
& \left.
{}+ 2 m^2 g_0^{-1} \bigl( 2 d_{16}^r - d_{18}^{} \bigr) 
  - 8 m^2 d_{28}^r
  + 4 m^2 \tilde{c}_n^{} \rule{0pt}{1.4em} \right)
+ \widetilde M_{n,n-1}^{I=1\slim (0)} + O(q)
\end{align}
with the low-energy constants $d_{16}^r(\mu)$, $d_{18}^{}$ and
$d_{28}^r(\mu)$ from \cite{Fettes:1998ud}.  To make the pion mass
dependence fully explicit, one should replace $M^2 = M_0^2 - 8 m^2 M_0
c_1 + O(m^3)$.  Conversely, we can use \eqref{Bn} and the one-loop
expression
\begin{equation}
g_A = g_0 \left( 1 - \frac{m^2}{(4\pi F)^2}
  \left[(2g_0^2+1) \log\frac{m^2}{\mu^2} + {g_0^2} \slim\right]
  + 4 m^2 g_0^{-1} d_{16}^r - 8 m^2 d_{28}^r \right) + O(m^3)
\end{equation}
of the axial coupling (see e.g.\ \cite{Bernard:2006te}) to rewrite the
result as
\begin{align}
  \label{Mnn-1-final}
\widetilde M_{n,n-1}^{I=1}(t) =
B_n^\pi\; \frac{4 M^2 g_A}{m_\pi^2 - t} \,
  \bigl( 1 - 2m_\pi^2\slim g_A^{-1} d_{18}^{} \bigr)
+ \widetilde M_{n,n-1}^{I=1\slim (0)} + O(q)
\end{align}
in terms of the physical quantities $B_n^\pi$, $g_A$, $M$, $m_\pi$ and
the low-energy constants $d_{18}$ and $\widetilde M_{n,n-1}^{I=1\slim
(0)}$.  With the transformation \eqref{ff-inverse} and the definitions
\eqref{gpd-mom} and \eqref{piDAmom} of moments one finds
\begin{equation}
\widetilde{E}^{I=1}(x,\xi,t) =
\frac{\theta(|x|<\xi)}{2\xi}\, \phi_\pi\Bigl( \frac{x}{\xi} \Bigr)\,
  \frac{4 M^2 g_A}{m_\pi^2 - t} \,
  \bigl( 1 - 2m_\pi^2\slim g_A^{-1} d_{18}^{} \bigr)
+ \widetilde{E}^{I=1\slim (0)}(x,\xi) + O(q) \,,
\end{equation}
which generalizes the well-known relation from
\cite{Mankiewicz:1998kg,Penttinen:1999th} to next-to-leading order in
the chiral expansion.  For $n=1$ our result \eqref{Mnn-1-final} is
consistent with the one for the pseudoscalar form factor in
\cite{BFHM,Bernard:2001rs}.  We also agree with the result of
Ref.~\cite{ando} if in their eq.~(66) one adds a term $\delta_{m,2k}\;
\frac{2}{3}\slim g_A M^2 \langle r_A^2 \rangle\, \langle
z^{2k}\rangle_\pi \big/ 2^{2k}$.

%%%%%%%%%%%%%%%%%%%%%%%%%%%%%%%%%%%%%%%%%%%%%%%%%%%%%%%%
\section{Chiral-odd generalized parton distributions}
\label{sect:tensor}
%%%%%%%%%%%%%%%%%%%%%%%%%%%%%%%%%%%%%%%%%%%%%%%%%%%%%%%%

In this section we consider the general parton distributions
associated with chiral-odd operators.  As in the previous sections we
restrict ourselves to the twist-two sector.  The relevant GPDs of the
nucleon are defined by\footnote{%
We have traded the distribution $E_T$ in the original decomposition
\protect\cite{Diehl:2001pm} for the combination $\widebar{E}_T = E_T +
2\widetilde{H}_T$, which naturally appears when representing the
distributions at $\xi=0$ in terms of densities in the impact parameter
plane \protect\cite{Diehl:2005jf}.}
\begin{align}
  \label{quark-tensor-gpd}
& \int \frac{d \eta}{4\pi}\, e^{ix \eta (aP)}
\bigl\langle N_i(p') \bigl|\, \bar{q}(-\half\eta a)\, 
  b_\lambda a_\mu\slim i\sigma^{\lambda\mu}\slim
  \tau^A q(\half\eta a) \,\bigr| N_j(p) \bigr\rangle
\nonumber \\
&\qquad = \tau^A_{ij}\,  \frac{1}{2 aP}\, b_\lambda a_\mu\,
\bar{u}(p') \Biggl[ i\sigma^{\lambda\mu} H_T^I
  + \frac{\gamma^\lambda \Delta^\mu- \Delta^\lambda \gamma^\mu}{2M}\,
     \widebar{E}_T^I\,
\nonumber\\[2mm]
&\hspace{11em}
  - \frac{i\sigma^{\lambda\nu} \Delta^\mu \Delta_\nu
        - i\sigma^{\mu\nu} \Delta^\lambda \Delta_\nu}{2 M^2}\,
    \widetilde H_T^I\,
  + \frac{\gamma^\lambda P^\mu- P^\lambda \gamma^\mu}{M}\, 
    \widetilde E_T^I
\Biggr]  u(p) \,.
\end{align}
In addition to the light-like vector $a$ we have introduced a vector
$b$ satisfying $ab=0$, and for brevity we have suppressed the
arguments $x$, $\xi$, $t$ of the distributions.\footnote{%
Instead of contracting $\sigma^{\lambda\mu}$ with auxiliary vectors
one often takes definite indices $\sigma^{i+}$, where $i=1,2$ denotes
a transverse component and $+$ the plus-component in light-cone
coordinates, i.e.\ $\sigma^{i+} = ( \sigma^{i0} + \sigma^{i3} )
/\sqrt{2}$.}
Their Mellin moments are related to the matrix elements of the
chiral-odd twist-two operators
\begin{equation}
  \label{tens-operators}
\mathcal{O}^{A}_{T\slim \lambda \mu_1 \mu_2 \ldots \mu_n} =
  \subtr{\lambda \mu_1 \ldots \mu_n} \; \Asy{\lambda \mu_1}
    \Sym{\phantom{\lambda} \mu_1 \ldots \mu_n}
  \bar{q} \sigma_{\lambda\mu_1}
       i \lrD_{\mu_2} \ldots i \lrD_{\mu_n} \tau^A q \,,
\end{equation}
where $\subtr{}$ and $\Sym{}$ are defined as in
Section~\ref{sect:even-gpds} and where $\Asy{}$ denotes
antisymmetrization, $\Asy{\lambda\mu} t^{\lambda\mu} = \half(
t^{\lambda\mu} - t^{\mu\lambda})$.  These operations are conveniently
implemented by contraction with the auxiliary vectors $a$ and $b$,
given that for any tensor $t^{\lambda\mu_1 \ldots \mu_n}$ satisfying
$t^{\lambda \mu_1 \mu_2 \ldots \mu_n} = - t^{\mu_1 \lambda \mu_2
\ldots \mu_n}$ one has
\begin{align}
  \label{contract-all}
 & b_\lambda a_{\mu_1} a_{\mu2} \ldots a_{\mu_n}
   \subtr{\lambda \mu_1 \ldots \mu_n} \; \Asy{\lambda \mu_1}
   \Sym{\phantom{\lambda} \mu_1 \ldots \mu_n}
      t^{\lambda\mu_1 \ldots \mu_n} 
\nonumber \\
 &= b_\lambda a_{\mu_1} a_{\mu2} \ldots a_{\mu_n} \frac{1}{2n}\,
    \Bigl( t^{\lambda\slim \mu_1 \mu_2 \ldots \mu_n}
    + \sum_{i=2}^{n} t^{\lambda\slim \mu_2 \ldots \mu_{i}\slim
               \mu_1\slim \mu_{i+1} \ldots \mu_n}
         - t^{\mu_1 \lambda\slim \mu_2 \ldots \mu_n}
    - \sum_{i=2}^{n} t^{\mu_1 \mu_2 \ldots \mu_{i}\slim
               \lambda\slim \mu_{i+1} \ldots \mu_n} \Bigr)
\nonumber \\
 &= \frac{n+1}{2n}\, b_{\lambda} a_{\mu_1} a_{\mu_2} \ldots a_{\mu_n}\,
    t^{\lambda\mu_1 \ldots \mu_n} \,.
\end{align}
where symmetrization in $\mu_2 \ldots \mu_n$ is guaranteed by
contraction with identical vectors, and where trace subtraction terms
are removed by the conditions $a^2 = ab =0$.  The $(n-1)$ terms of the
second sum give zero due to the antisymmetry of $t$ in its first two
indices.  We therefore define the contracted operator
\begin{equation}
  \label{contract-ops}
\mathcal{O}^{A}_{T n}(a,b) = \frac{2n}{n+1}\,
   b^{\lambda} a^{\mu_1} a^{\mu_2} \ldots a^{\mu_n}\,
   \mathcal{O}^{A}_{T\slim \lambda \mu_1 \mu_2 \ldots \mu_n}
 = \bar{q}\, b^\lambda a^\mu\slim i\sigma_{\lambda\mu}\slim 
   (ia\lrD)^{n-1} \tau^A q \,,
\end{equation}
whose nucleon matrix elements are parameterized by
\begin{align}
  \label{quark-tensor-ffs}
\bigl\langle N_i(p') \bigl|\, \mathcal{O}^{A}_{T n}(a,b)
  \,\bigr| N_j(p) \bigr\rangle
 &= \tau^A_{ij}\,  \sum_{k=0}^{n-1}
   (a P)^{n-k-1}\, (a  \Delta)^k\; {b_\lambda a_\mu}\, 
\bar{u}(p') \Biggl[ i\sigma^{\lambda\mu} A^{I}_{T n,k} +
  \frac{\gamma^\lambda \Delta^\mu- \Delta^\lambda \gamma^\mu}{2M}\, 
     \widebar{B}^{I}_{T n,k}\,
\nonumber\\[0.5em]
&\quad
  - \frac{i\sigma^{\lambda\nu} \Delta^\mu \Delta_\nu
        - i\sigma^{\mu\nu} \Delta^\lambda \Delta_\nu}{2 M^2}\,
    \widetilde A^{I}_{T n,k}\,
  + \frac{\gamma^\lambda P^\mu- P^\lambda \gamma^\mu}{M}\, 
    \widetilde B^{I}_{T n,k}
\Biggr]  u(p) \,.
\end{align}
The moments of the chiral-odd GPDs are then expressed as
\cite{Hagler:2004yt} 
\begin{align}
  \label{gpd-mom-t}
\int_{-1}^1dx\, x^{n-1}\, H_T(x,\xi,t) &=
\sum_{\substack{k=0\\{\mathrm{even}}}}^{n-1}
(-2\xi)^k\, A_{T n,k}(t) \,,
&
\int_{-1}^1dx\, x^{n-1}\, \widebar{E}_T(x,\xi,t) &=
\sum_{\substack{k=0\\{\mathrm{even}}}}^{n-1}
(-2\xi)^k\, \widebar{B}_{T n,k}(t) \,,
\nonumber \\
\int_{-1}^1dx\, x^{n-1}\, \widetilde{H}_T(x,\xi,t) &=
\sum_{\substack{k=0\\{\mathrm{even}}}}^{n-1}
(-2\xi)^k\, \widetilde{A}_{T n,k}(t) \,,
&
\int_{-1}^1dx\, x^{n-1}\, \widetilde{E}_T(x,\xi,t) &=
\sum_{\substack{k=1\\{\mathrm{odd}}}}^{n-1}
(-2\xi)^k\, \widetilde{B}_{T n,k}(t) \,,
\end{align}
where we have omitted isospin indices $I$ in the distributions and
form factors for ease of writing.  The restrictions to even or odd $k$
for the form factors reflect that $H_T$, $\widebar{E}_T$ and
$\widetilde{H}_T$ are even in $\xi$ and $\widetilde{E}_T$ is odd in
$\xi$ due of time reversal invariance \cite{Diehl:2001pm}.

Using the relations (9) in \cite{DMS2} and
\begin{equation}
  \label{sred}
\bar u(p')i\sigma^{\lambda\mu}u(p) =
\bar{u}_v(p')\, \Biggl[\slim
2\gamma\slim [S^\lambda,S^\mu] 
+ \frac{v^\lambda \Delta^\mu - v^\mu \Delta^\lambda}{2M}\,
+ \frac{[S^\lambda,(S\Delta)] \Delta^\mu 
      - [S^\mu,(S\Delta)] \Delta^\lambda}{2M^2(1+\gamma)}
\slim\Biggr] \,  u_v(p)
\end{equation}
one can rewrite the decomposition \eqref{quark-tensor-ffs} in terms of
the heavy-baryon spinors \eqref{spinors} and obtains
\begin{align}
  \label{tensor-ffs}
& \bigl\langle N_i(p') \bigl|\, \mathcal{O}^{A}_{T n}(a,b)
  \,\bigr| N_j(p) \bigr\rangle
 = \tau^A_{ij}\,  \sum_{k=0}^{n-1} (M\gamma)^{n-k-1}
   (a v)^{n-k-1}\, (a  \Delta)^k\; {b_\lambda a_\mu}\, 
\nonumber\\
& \hspace{1em} \quad\times
\bar u_v(p')\,\Biggl[\slim
2 \gamma\slim [S^\lambda, S^\mu]\, \widetilde{E}^{I}_{T n,k}
+ \frac{[S^\lambda, (S\Delta)]\slim \Delta^\mu 
      - [S^\mu, (S\Delta)]\slim \Delta^\lambda}{2M^2}\,
\widetilde{M}^{I}_{T n,k}
\nonumber\\[2mm]
& \hspace{5.8em} 
+ \gamma\, \frac{[S^\lambda, (S\Delta)]\slim v^\mu 
               - [S^\mu, (S\Delta)]\slim v^\lambda}{M}\, 
  E^{I}_{T n,k}
+ \frac{v^\lambda \Delta^\mu - v^\mu \Delta^\lambda}{2M}\,
  M^{I}_{T n,k}
\Biggr] \, u_v(p)
\end{align}
with new form factors given by
\begin{align}
  \label{ff-trafo-odd}
\widetilde{E}^{}_{T n,k} &= A^{}_{T n,k} \,,
&
\widetilde{M}^{}_{T n,k} &= 
   (1+\gamma)^{-1} A^{}_{T n,k}
   + \widebar{B}^{}_{T n,k} - 2 \widetilde{A}^{}_{T n,k} \,.
\nonumber \\[0.4em]
E^{}_{T n,k} &= \widetilde{B}^{}_{T n,k} \,,
&
M^{}_{T n,k} &= A^{}_{T n,k}
  + \widebar{B}^{}_{T n,k}
  - \frac{\Delta^2}{2M^2}\, \widetilde{A}^{}_{T n,k} \,,
\end{align}
or equivalently
\begin{align}
\widebar{B}^{}_{T n,k} &=
\frac{1}{\gamma^2} \left[ M^{}_{T n,k}
  - \frac{\Delta^2}{4M^2}\, \widetilde{M}^{}_{T n,k}
  - \gamma\slim \widetilde{E}^{}_{T n,k} \right] \,,
\nonumber \\
\widetilde{A}^{}_{T n,k} &=
\frac{1}{2\gamma^2} \left[ M^{}_{T n,k} - \widetilde{M}^{}_{T n,k} 
  - \frac{\gamma}{1+\gamma}\, \widetilde{E}^{}_{T n,k} \right] \,.
\end{align}

We finish this section by defining chiral-odd GPDs in the pion,
\begin{align}
  \label{pion-tensor-gpd}
& \int \frac{d \eta}{4\pi}\, e^{ix \eta (aP)}
\bigl\langle \pi^c(p') \bigl|\, \bar{q}(-\half\eta a)\, 
  b_\lambda a_\mu\slim i\sigma^{\lambda\mu}\slim
  \tau^A q(\half\eta a) \,\bigr| \pi^b(p) \bigr\rangle
\nonumber \\
&\qquad = \frac{1}{2} \Tr(\tau^A \tau^b \tau^c)\;
  \frac{1}{2 aP}\, b_\lambda a_\mu\,
  \frac{P^\lambda \Delta^\mu - \Delta^\lambda P^\mu}{m_\pi}\,
  E_{T \pi}^{I}(x,\xi,t) \,,
\end{align}
where as in the nucleon case, isospin $I=0$ corresponds to $A=0$ and
$I=1$ to $A=1,2,3$.  In terms of quark flavors in a $\pi^+$ one has
$E_{T \pi}^{I=0} = E_{T \pi}^{u} + E_{T ,\pi}^{d}$ and $E_{T \pi}^{I=1}
= E_{T \pi}^{u} - E_{T \pi}^{d}$, with the definition
\begin{align}
& \int \frac{d \eta}{4\pi}\, e^{ix \eta (aP)}
\bigl\langle \pi^+(p') \bigl|\, \bar{u}(-\half\eta a)\, 
  b_\lambda a_\mu\slim i\sigma^{\lambda\mu}\slim
  u(\half\eta a) \,\bigr| \pi^+(p) \bigr\rangle
\nonumber \\
&\qquad = \frac{1}{2 aP}\, b_\lambda a_\mu\,
  \frac{P^\lambda \Delta^\mu - \Delta^\lambda P^\mu}{m_\pi}\,
  E_{T \pi}^{u}(x,\xi,t)
\end{align}
and its analog for $d$ quarks.  For the local twist-two operators one
has 
\begin{equation}
  \label{pion-tensor-ffs}
\bigl\langle \pi^c(p') \bigl|\, \mathcal{O}^{A}_{T n}(a,b)
  \,\bigr| \pi^b(p) \bigr\rangle
= \frac{1}{2} \Tr(\tau^A \tau^b \tau^c)\;
  b_\lambda a_\mu\, 
  \frac{P^\lambda \Delta^\mu - \Delta^\lambda P^\mu}{m_\pi}\,
  \sum_{\substack{k=0\\{\mathrm{even}}}}^{n-1}
    (a P)^{n-k-1}\, (a\Delta)^{k}\, B^{\pi}_{T n,k}(t)
\end{equation}
with
\begin{equation}
\int_{-1}^1dx\, x^{n-1}\, E^I_{T \pi}(x,\xi,t) =
\sum_{\substack{k=0\\{\mathrm{even}}}}^{n-1}
(2\xi)^k\, B^{\pi}_{T n,k}(t) \,,
\end{equation}
where the restriction to even $k$ is a consequence of time reversal
symmetry.  Due to isospin and charge conjugation invariance, $n$ is
even for $I=0$ and odd for $I=1$, so that we do not need an isospin
label for $B^{T \pi}_{n,k}$.

%%%%%%%%%%%%%%%%%%%%%%%%%%%%%%%%%%%%%%%%%%%%%%%%%%%%%%%%
\section{Chiral-odd effective operators}
\label{sect:tensor-ops}
%%%%%%%%%%%%%%%%%%%%%%%%%%%%%%%%%%%%%%%%%%%%%%%%%%%%%%%%

In this section we explain how to construct the operators in the
effective theory that match the chiral-odd quark operators
\eqref{tens-operators}, closely following the strategy used in
Section~\ref{sect:isotriplet}.  To this end we first match the
operators 
\begin{align}
  \label{proto-ops}
\bigl( \mathcal{O}^{\lambda\mu}_{RL, n}(a) \bigr)_{ij} &=
  \bar{q}_j\slim i\sigma^{\lambda\mu}\, 
      \frac{1+\gamma_5}{2}\, (i a\lrD)^{n-1}\, q_i \,,
&
\bigl( \mathcal{O}^{\lambda\mu}_{LR, n}(a) \bigr)_{ij} &=
  \bar{q}_j\slim i\sigma^{\lambda\mu}\, 
      \frac{1-\gamma_5}{2}\, (i a\lrD)^{n-1}\, q_i
\end{align}
with open isospin indices $i$, $j$, which involve quarks of definite
chirality.  The corresponding uncontracted operators $\half
\bar{q}_j\slim i\sigma_{\lambda \mu_1} (1\pm \gamma_5)\, i
\lrD_{\mu_2} \ldots i \lrD_{\mu_n}\, q_i$ do not have definite twist,
but according to \eqref{contract-all} their twist-two part is readily
projected out in $b_\lambda a_\mu\, \mathcal{O}^{\lambda\mu}_{RL,
n}(a)$ and $b_\lambda a_\mu\, \mathcal{O}^{\lambda\mu}_{LR, n}(a)$.
The operators \eqref{proto-ops} transform as
\begin{align}
  \label{RL-trans}
\mathcal{O}^{\lambda\mu}_{RL, n}(a) &\to
   V_R^{\phantom{\dagger}}\slim \mathcal{O}^{\lambda\mu}_{RL, n}(a)
   \slim V_L^\dagger
&
\mathcal{O}^{\lambda\mu}_{LR, n}(a) &\to
   V_L^{\phantom{\dagger}}\slim \mathcal{O}^{\lambda\mu}_{LR, n}(a)
   \slim V_R^\dagger
\end{align}
unter chiral rotations and are transformed into each other by parity.
Because $\sigma^{\lambda\mu}\gamma_5 = -\half i\slim
\epsilon^{\lambda\mu}{}_{\alpha\beta}\, \sigma^{\alpha\beta}$ they
obey the duality relations
\begin{align}
  \label{duality}
\mathcal{O}_{RL, n}^{\lambda\mu}(a) &= -\frac{i}{2}\slim
  \epsilon^{\lambda\mu}{}_{\alpha\beta}\,
  \mathcal{O}_{RL, n}^{\alpha\beta}(a)
&
\mathcal{O}_{LR, n}^{\lambda\mu}(a) &= \frac{i}{2}\slim
  \epsilon^{\lambda\mu}{}_{\alpha\beta}\,
  \mathcal{O}_{LR, n}^{\alpha\beta}(a)
\end{align}
The operators $\mathcal{O}^{A}_{T n}(a,b)$ from \eqref{contract-ops},
which correspond to twist-two and to definite isospin, are obtained as
\begin{align}
\mathcal{O}^{A}_{T n}(a,b) &= 
   b_\lambda a_\mu\, Q^{A, \lambda\mu}_{n}(a) \,,
&
Q^{A, \lambda\mu}_{n}(a) &= \Tr \tau^A\slim 
  \bigl\{ \mathcal{O}_{RL, n}^{\lambda\mu}(a) 
        + \mathcal{O}_{LR, n}^{\lambda\mu}(a) \bigr\} \,.
\end{align}
They will involve the combinations
\begin{equation}
  \label{tau-odd-def}
\tau^A_{\,o \pm} = u^\dagger \tau^A u^\dagger \pm u \tau^A u \,,
\end{equation}
whose expansion in pion fields reads
\begin{align}
  \label{tau-odd}
\tau^0_{o+} &= 2 - \frac{1}{F^2}\, \pi^a \pi^a + O(\pi^4) \,,
&
\tau^0_{o-} &= - \frac{2i}{F}\, \pi^a \tau^a + O(\pi^3) \,,
\nonumber \\[0.1em]
\tau^a_{o+} &= 2\tau^a - \frac{1}{F^2}\, \pi^a\pi^b\tau^b + O(\pi^4) \,,
&
\tau^a_{o-} &= - \frac{2i}{F}\, \pi^a + O(\pi^3) \,.
\end{align}
As for the chiral-even case discussed in Section~\ref{sect:operators},
the operators which match \eqref{proto-ops} in the effective theory
and contribute to nucleon matrix elements are either bilinear in the
nucleon field or contain only pion operators.  We treat the two cases
in the following two subsections.

%%%%%%%%%%%%%%%%%%%%%%%%%%%%%%%%%%%%%%%%%%%%%%%%
\subsection{Pion-nucleon operators}
\label{sect:piN}
%%%%%%%%%%%%%%%%%%%%%%%%%%%%%%%%%%%%%%%%%%%%%%%%

The effective operators which are bilinear in the nucleon field and
transform as \eqref{RL-trans} can be written in the form
\begin{align}
\bigl( \mathcal{O}^{\lambda\mu}_{RL, n}(a) \bigr)_{ij}
 &= \bigl( \Nvbar\slim \mathcal{O}_1 u \bigr)_{j}\,
    \bigl( u\slim \mathcal{O}_2 N_v \bigr)_{i} \,,
&
\bigl( \mathcal{O}^{\lambda\mu}_{LR, n}(a) \bigr)_{ij}
 &= \bigl( \Nvbar\slim \mathcal{O}'_1 u^\dagger \bigr)_{j}\,
    \bigl( u^\dagger\slim \mathcal{O}'_2 N_v \bigr)_{i} \,,
\end{align}
where $\mathcal{O}_{1}$, $\mathcal{O}_{2}$ involve the fields $u_\mu$,
$\chi_\pm$ and covariant derivatives and transform like $u_\mu$ under
chiral rotations.  $\mathcal{O}'_{1}$ and $\mathcal{O}'_{2}$ are
related to $\mathcal{O}_{1}$ and $\mathcal{O}_{2}$ by parity.  One can
rearrange the covariant derivatives in $\mathcal{O}^{\lambda\mu}_{RL,
n}$ and $\mathcal{O}^{\lambda\mu}_{LR, n}$ such that they act either
as total derivatives $\partial_\mu$ or in the antisymmetric form
$\lrnab_\mu = \half (\rnab_\mu - \lnab_\mu)$.

To obtain the general form of $\mathcal{O}^{\lambda\mu}_{RL, n}$ and
$\mathcal{O}^{\lambda\mu}_{LR, n}$ it is sufficient to construct
corresponding operators $\mathcal{O}^{\lambda\mu}$ that involve no
$\epsilon$ tensor and either no spin vector or two spin vectors in the
form $[S_\lambda, S_\mu]$.  Operators with one $\epsilon$ tensor and
one spin vector can be brought into this form by using the third
relation in \eqref{SS} and rewriting the resulting product of two
$\epsilon$ tensors in terms of products of metric tensors.  Terms in
$\mathcal{O}^{\lambda\mu}_{RL, n}$ and $\mathcal{O}^{\lambda\mu}_{LR,
n}$ with an odd total number of $\epsilon$ tensors and spin vectors
are then readily obtained by adding the dual operators $\half i\slim
\epsilon^{\lambda\mu}{}_{\alpha\beta}\, \mathcal{O}^{\alpha\beta}$
with coefficients determined by the relations~\eqref{duality}, using
that $\half i\slim \epsilon^{\lambda\mu}{}_{\alpha\beta}\, \half
i\slim \epsilon^{\alpha\beta}{}_{\gamma\delta}\, t^{\gamma\delta} =
t^{\lambda\mu}$ for any antisymmetric tensor $t^{\lambda\mu}$.

Following the procedure of Section~\ref{sect:operators} we decompose
the pion-nucleon part of the operators $Q^{A, \lambda\mu}_{n}(a)$ as
\begin{equation}
  \label{tensor-decom}
Q^{\lambda\mu}_{n, \pi N}(a) = 
\sum_{k=0}^{n-1} M^{n-k-1} (av)^{n-k-1}\, Q^{\lambda\mu}_{n,k}(a) \,,
\end{equation}
where we have omitted superscripts $A$ for ease of writing.  The
operator $Q^{\lambda\mu}_{n,k}(a)$ is the contraction of a tensor of
rank $k+2$ with $k$ vectors $a$ and may not contain any factors
$(av)$.  The minimal number of vectors $\partial_\rho$, $\lrnab_\rho$,
$u_\rho$ in $Q^{\lambda \mu}_{n,k}(a)$ is $k-1$ and must be
accompanied either by the tensor $v^\lambda [S^\mu, (aS)] - v^\mu
[S^\lambda, (aS)]$ or by its dual $i
\epsilon^{\lambda\mu\alpha\beta}\, v_\alpha [S_\beta, (aS)]$.  In the
first case one obtains however the structure $(av) [(bS), (aS)]$ after
contraction with $b_\lambda a_\mu$, which also appears in $b_\lambda
a_\mu\, (av)^{n-k}\, Q^{\lambda\mu}_{n,k-1}(a)$.  An analogous
statement holds of course in the case of the dual tensor.  We can
therefore restrict ourselves to operators $Q^{\lambda\mu}_{n,k}(a)$
with at least $k$ vectors $\partial_\rho$, $\lrnab_\rho$, $u_\rho$,
and thus further decompose
\begin{equation}
  \label{tensor-dec}
Q^{\lambda\mu}_{n,k}(a) = \sum_{i=0}^\infty
  M^{-i}\, Q^{\lambda\mu}_{n,k,i}(a) \,,
\end{equation}
where $Q^{\lambda\mu}_{n,k,i}(a)$ has chiral dimension $k+i$.  The
power counting for graphs with a certain operator insertion proceeds
in close analogy to Section~\ref{sect:power} and is summarized in
Table~\ref{tab:tensor-order}.  Comparing the number of factors $(av)$
in \eqref{tensor-decom} and in the decomposition \eqref{tensor-ffs},
one obtains the restriction $l\ge k$ for the operators
$Q^{\lambda\mu}_{n,l,i}(a)$ that can contribute to $\widetilde{E}_{T
n,k}$ and $\widetilde{M}_{T n,k}$.  For $E_{T n,k}$ and $M_{T n,k}$
the restriction is $l\ge k-1$, where the case $l=k-1$ requires that
the graphs with insertion of $\smash{Q^{\lambda\mu}_{n,l,i}(a)}$
produce no factors of $v^\lambda$ or $v^\mu$.

\begin{table}
\caption{\label{tab:tensor-order} Overview of contributions to the
  chiral-odd form factors.  As in Table~\protect\ref{tab:chiral-order}
  the restriction in the second column is due to time reversal
  invariance.  $N_\Delta$ is the number of factors $(a\Delta)$,
  $(b\Delta)$ and $(S\Delta)$ in the decomposition
  \protect\eqref{tensor-ffs}.  One must have $l\ge k-1$ for $E_{T
  n,k}$, $M_{T n,k}$ and $l\ge k$ for $\widetilde{E}_{T n,k}$,
  $\widetilde{M}_{T n,k}$, and $i\ge 0$ for all cases.  The
  corresponding graphs contribute to the form factor at order
  $O(q^{\slim d})$ with $d \ge D_{l,i} - N_\Delta$ and $D_{l,i}$ from
  \protect\eqref{pow-count}.}
\begin{center}
  \renewcommand{\arraystretch}{1.3}
  \begin{tabular}{cccc} \hline\hline
    form factor & $k$  & $N_\Delta$ & operators \\ \hline
    $E_{T n,k}$ & odd  & $k+1$ & $Q^{\lambda\mu}_{n,l,i}$ \\
    $M_{T n,k}$ & even & $k+1$ & $Q^{\lambda\mu}_{n,l,i}$ \\
    $\widetilde{E}_{T n,k}$ & even & $k$   & $Q^{\lambda\mu}_{n,l,i}$ \\
    $\widetilde{M}_{T n,k}$ & even & $k+2$ & $Q^{\lambda\mu}_{n,l,i}$ 
    \\[0.2em]
    \hline\hline
  \end{tabular}
\end{center}
\end{table}

%%%%%%%%%%%%%%%%%%%%%%%%%%%%%%%%%%%%%%%%%%%%%%%%
\subsection{Pure pion operators}
\label{sect:pure-pi}
%%%%%%%%%%%%%%%%%%%%%%%%%%%%%%%%%%%%%%%%%%%%%%%%

Pionic operators which transform according to \eqref{RL-trans} can be
written as
\begin{align}
\mathcal{O}^{\lambda\mu}_{RL, n}(a) 
 &= u\slim \mathcal{O} u \,,
&
\mathcal{O}^{\lambda\mu}_{LR, n}(a)
 &= u^\dagger\slim \mathcal{O}'\slim u^\dagger \,,
\end{align}
where $\mathcal{O}$ and $\mathcal{O}'$ are related by a parity
transformation, transform like $u_\mu$ under chiral rotations, and are
constructed from the fields $u_\mu$, $\chi_\pm$ and covariant
derivatives.  We can restrict the derivatives to act only on fields
inside $\mathcal{O}$ and $\mathcal{O}'$.\footnote{%
Other terms can be brought into this form using identities such as
$\partial_\rho\slim (u\slim \mathcal{O} u) = u\slim \bigl( \slim
[\nabla_\rho, \mathcal{O}\slim ] - \frac{i}{2} \{u_\rho , \mathcal{O}
\} \slim\bigr)\slim u$.}
With the duality relations \eqref{duality} one finds that the pure
pion part of the operator $Q^{A, \lambda\mu}_{n}(a)$ can be brought
into the form
\begin{align}
  \label{tensor-decom-pi}
\Tr \Bigl[ \tau_{o+}^A V^{\lambda\mu}(a) \Bigr]
  + \frac{i}{2}\slim \epsilon^{\lambda\mu\alpha\beta} 
    \Tr \Bigl[ \tau_{o-}^A V_{\alpha\beta}(a) \Bigr]
& & \text{or} & &
\Tr \Bigl[ \tau_{o-}^A\slim A^{\lambda\mu}(a) \Bigr]
  + \frac{i}{2}\slim \epsilon^{\lambda\mu\alpha\beta} 
    \Tr \Bigl[ \tau_{o+}^A\slim A_{\alpha\beta}(a) \Bigr] \,,
\end{align}
where $V_{\lambda\mu}$ and $A_{\lambda\mu}$ respectively behave as a
tensor or a pseudotensor under parity and are constructed from
$u_\rho$, $\nabla_\rho$ and $\chi_{\pm}$ without the $\epsilon$
tensor.  One readily finds that the terms without $\epsilon$ in
\eqref{tensor-decom-pi} couple to an even number and the terms with
$\epsilon$ to an odd number of pion fields.  $V_{\lambda\mu}(a)$ and
$A_{\lambda\mu}(a)$ are tensors of rank $n+1$ contracted with $n-1$
vectors $a$.  In the following we consider the terms with the lowest
chiral dimension in the pure pion part of $\mathcal{O}^{A}_{T
n}(a,b)$.  These terms contain no fields $\chi_{\pm}$ and have the
vector indices of all $n+1$ factors $u_\rho$ or $\nabla_\rho$
contracted with either $a$ or $b$.

To calculate matrix elements of these operators between two nucleons
or two pions at one-loop accuracy, we only need terms that couple to
at most four pions.  Terms coupling to three or four pions can appear
in tadpole graphs.  Such graphs are only nonzero if the pion fields
in the operator which couple to the loop have no derivatives acting on
them.  This is because the corresponding loop integral has a numerator
of the form $l_{\rho_1} \ldots l_{\rho_m}$, where $l$ is the loop
momentum.  After the loop integration, one obtains zero for odd $m$
and for even $m$ one obtains a combination of metric tensors, which
gives zero when the vector indices are contracted with $a$ or $b$.

Since the derivatives with indices $\lambda$ and $\mu$ in the
antisymmetric tensor $Q^{A, \lambda\mu}_{n}(a)$ cannot act on the same
pion field, one readily finds that operators coupling to one or three
pions do not contribute to matrix elements between two nucleons or two
pions.  For the same reason such operators decouple from matrix
elements between the vacuum and a single pion, which reflects the fact
that there are no chiral-odd pion distribution amplitudes of twist two.

It remains to construct operators $V_{\lambda\mu}(a)$ and
$A_{\lambda\mu}(a)$ from $u_\rho$ and $\nabla_\rho$, which must have
at least one factor $u_\rho$ because the covariant derivatives must
act on some field to give nonzero, and less than three such factors
because of the restriction just discussed.  The operators with one
factor of $u_\rho$ are of the form\footnote{%
For simplicity we write from now on $\nabla_\mu \mathcal{O}$ instead
of $[\nabla_\mu, \mathcal{O} \slim]$ if $\mathcal{O}$ transforms like
$u_\mu$ under chiral rotations.}
\begin{align}
(a\nabla)^{k_1}\, \nabla^\lambda\, (a\nabla)^{k_2}\, u^\mu
  - (\lambda\leftrightarrow\mu) 
& & \text{or} & & 
(a\nabla)^{k_1}\, \nabla^\lambda\, (a\nabla)^{k_2}\, \nabla^\mu\,
    (a\nabla)^{k_3}\, (au) - (\lambda\leftrightarrow\mu)  \,.
\end{align}
In both cases we can use the commutator identity $[\nabla^\alpha,
\nabla^\beta]\, \mathcal{O} = \frac{1}{4} \bigl( [u^\alpha, u^\beta]\,
\mathcal{O} - \mathcal{O}\, [u^\alpha, u^\beta] \bigr)$ to bring the
vectors with indices $\lambda$ and $\mu$ next to each other.  The
commutator terms do not contribute to the matrix element in question
since they involve three or more vectors $u_\rho$.  The remaining term
involves either $\nabla^\lambda u^\mu - \nabla^\mu u^\lambda = 0$ or
$[\nabla^\lambda, \nabla^\mu] \ldots (au)$ and thus do not contribute
either.  

The only relevant operators contain hence two vectors $u_\rho$.
According to our above discussion, the $\Gamma_\rho$ part of any
factor $\nabla_\rho$ does not contribute in this case, and the
derivative must act on the pion fields in $u_\rho$ which already carry
a derivative.  For the matrix elements in question, $\nabla^\lambda\,
(au)$ is hence equivalent to $(a\nabla)\, u^\lambda$.  The same holds
of course for the index $\mu$.  We thus find that the operators of
lowest chiral dimension can be written as
\begin{align}
  \label{pion-tensor-ops}
Q^{A, \lambda\mu}_{n, \pi}(a) &= \frac{F^2}{8}\,
 \sum_{\substack{k=0 \\ \mathrm{even}}}^{n-1} b^{}_{T n,k}\,
   \Tr \tau^{A}_{o+}\slim (ia\nabla)^k\, V_{n,k}^{\lambda\mu} + \ldots
\\
\intertext{with}
V_{n,k}^{\lambda\mu} &= u^{\lambda}\, (2ia\lrnab)^{n-k-1}\, u^{\mu}
                      - u^{\mu}\, (2ia\lrnab)^{n-k-1}\, u^{\lambda} \,,
\end{align}
where the $\ldots$ denote terms not contributing to two-nucleon or
two-pion matrix elements at tree level or one loop.  We note that the
coefficients $b_{T n,k}$ have nonzero mass dimension and are of order
$(4\pi F)^{-1}$ in the sense of chiral power counting.  They give the
tree-level contribution at order $O(q^0)$ to the pion form factors
$B^{\pi}_{T n,k}(t)$ defined in \eqref{pion-tensor-ffs},
\begin{equation}
  B^{\pi\slim (0)}_{T n,k} =
     (-1)^{n+1}\, 2^{n-k-1}\, m_\pi\slim b^{}_{T n,k} ,
\end{equation}
where $n$ is even in the isosinglet and odd in the isotriplet case.
The restriction to even $k$ in \eqref{pion-tensor-ops} corresponds to
the one in \eqref{pion-tensor-ffs}.

We can now apply the power counting formula \eqref{pion-dim} with
$d_\pi = n+1$ to the operators just constructed.  Taking into account
the restrictions of even or odd $n$ or $k$ for the different form
factors, we find that the corrections from pion operator insertions
start at order $O(q)$ for $\widetilde{M}_{T n,n-1}^{I=1}$ and at order
$O(q^2)$ for $M_{T n,n-1}^{I=1}$, $E_{T n,n-1}^{I=0}$ and
$\widetilde{M}_{T n,n-2}^{I=0}$.  For all other form factors they
start at order $O(q^3)$ or higher.

%%%%%%%%%%%%%%%%%%%%%%%%%%%%%%%%%%%%%%%%%%%%%%%%%%%%%%%%%%%%%%
\section{Results for chiral-odd form factors} 
\label{sect:tensor-results}
%%%%%%%%%%%%%%%%%%%%%%%%%%%%%%%%%%%%%%%%%%%%%%%%%%%%%%%%%%%%%%

Using the construction described in Section~\ref{sect:tensor-ops}, we
find pion-nucleon operators
\begin{align}
  \label{piN-ops}
Q_{n,k,0}^{A, \lambda\mu} &=
  \widetilde{E}^{I\slim (0)}_{T n,k}\, (ia\partial)^k\,
  \Bigl( \, \Nvbar [S^\lambda, S^\mu]\slim \tau^A_{\,o +} N_v
     + \Nvbar \bigl( v^\lambda S^\mu - v^\mu S^\lambda \bigr) 
       \tau^A_{\,o -} N_v \Bigr) + \ldots \,,
\nonumber \\[0.3em]
Q_{n,k,1}^{A, \lambda\mu} &=
  - \frac{i}{4}\slim M^{I\slim (0)}_{T n,k}\, (ia\partial)^k\, \Bigl(
    \bigl( v^\lambda \partial^\mu - v^\mu \partial^\lambda \bigr)\,
      \Nvbar \tau^A_{\,o +} N_v 
    + i\epsilon^{\lambda\mu\alpha\beta} v_\alpha \partial_\beta 
      \Nvbar \tau^A_{\,o -} N_v \Bigr) + \ldots \,,
\nonumber  \\[0.3em]
& \quad + \frac{i}{2}\slim E^{I\slim (0)}_{T n,k}\, (ia\partial)^k\,
  \partial_\rho \Bigl( 
       v^\mu\slim \Nvbar [S^\lambda,S^\rho]\slim \tau^A_{\,o +} N_v
     - v^\lambda\slim \Nvbar [S^\mu,S^\rho]\slim \tau^A_{\,o +} N_v
  + \bigl\{ \text{terms with $\tau^A_{\,o -}$} \bigr\} \Bigr) 
+ \ldots \,,
\nonumber \\[0.3em]
Q_{n,k,2}^{A, \lambda\mu} &= 
  - \frac{1}{4}\slim
    \widetilde{M}^{I\slim (0)}_{T n,k}\, (ia\partial)^k\,
  \partial_\rho \Bigl(
      \partial^\mu\slim \Nvbar [S^\lambda, S^\rho]\slim 
         \tau^A_{\,o +} N_v
    - \partial^\lambda\slim \Nvbar [S^\mu, S^\rho]\slim 
         \tau^A_{\,o +} N_v
  + \bigl\{ \text{terms with $\tau^A_{\,o -}$} \bigr\} \Bigr) 
+ \ldots \,,
\end{align}
where the $\ldots$ denote terms with a smaller number of total
derivatives.  The coefficients in \eqref{piN-ops} are the tree-level
contributions at order $O(q^0)$ to the respective form factors and
therefore only nonzero for even or odd $k$ as given in
Table~\ref{tab:tensor-order}.  The terms with $\tau^A_{\,o -}$ in the
last two lines of \eqref{piN-ops} are rather lengthy and not given
here.  Indeed, one finds that none of the operators with $\tau^A_{\,o
-}$ in \eqref{piN-ops} contributes in one-loop graphs with
pion-nucleon interactions at LO or at NLO.  Such graphs have the form
of Fig.~\ref{fig-1}c and give zero for the same reasons discussed
after \eqref{zero-loop} for the case of operators with $\tau^A_{\,e
-}$.  The discussion at the end of Section~\ref{sect:ops} also applies
to the operators with $\tau^A_{\,o +}$ in \eqref{piN-ops}, so that
their insertion into graphs with LO pion-nucleon vertices or with one
NLO pion-nucleon interaction contributes only to those form factors
for which they already provide the tree-level result at order
$O(q^0)$.

An operator $Q_{n,l,i}^{\lambda\mu}$ in \eqref{piN-ops} has at most
$l+i$ partial derivatives, so that the condition \eqref{delta-cond}
holds also in the chiral-odd case.  Together with the power counting
following from Table~\ref{tab:tensor-order}, one again finds that
one-loop corrections to all form factors start at order $O(q^2)$.  One
finds that the order $O(q^2)$ corrections to $M_{T n,k}$ and $E_{T
n,k}$ come from $Q_{n,k,1}$, whereas those to $\widetilde{E}_{T n,k}$
and $\widetilde{M}_{T n,k}$ come from $Q_{n,k,0}$ and $Q_{n,k,2}$,
respectively, with pion-nucleon interactions taken at LO in all cases.
Additional order $O(q^2)$ contributions to $\widetilde{M}_{T n,k}$
come from graphs with $Q_{n,k,0}$ and two pion-nucleon interactions at
NLO or one pion-nucleon interaction at NNLO (only the $\tau^A_{\,o +}$
part of the operator is found to contribute).  Contributions from the
same graphs to $E_{T n,k+1}$ or $M_{T n,k+1}$ are possible by power
counting but turn out to be zero.  Other contributions at order
$O(q^2)$ which are possible by power counting involve at most one
pion-nucleon interaction at NLO and do not appear for the reason given
at the end of the preceding paragraph: there is no correction to $M_{T
n,k}$ or $E_{T n,k}$ from $Q_{n,k+1,0}$, $Q_{n,k,0}$, $Q_{n,k-1,1}$ or
$Q_{n,k-1,2}$ and no correction to $\smash{\widetilde{M}_{T n,k}}$
from $Q_{n,k+1,0}$, $Q_{n,k+2,0}$, $Q_{n,k,1}$ or $Q_{n,k+1,1}$.

Taking into account the graphs with pion operator insertions shown in
Fig.~\ref{fig-2}a and b, we finally find
\begin{align}
  \label{MI}
M_{T n,k}^{I=0 \phantom{()}} &=
M_{T n,k}^{I=0\slim (0)}
\left( 1 - \frac{3m^2}{2 (4\pi F)^2}\, \log\frac{m^2}{\mu^2} \right)
+ \ldots \,,
\nonumber \\[0.2em]
M_{T n,k}^{I=1 \phantom{()}} &=
M_{T n,k}^{I=1\slim (0)}
\left( 1 - \frac{m^2}{2 (4\pi F)^2}\,
\left[ \bigl(6g_A^2+1\bigr) \log\frac{m^2}{\mu^2} 
  + 4 g_A^2 \right] \right) 
+ \delta_{k,n-1}^{\phantom{()}}\, M_{T n,n-1}^{I=1\slim (2,\pi)}(t)
+ \ldots \,,
\nonumber \\[0.4em]
E_{T n,k}^{I=0 \phantom{()}} &=
E_{T n,k}^{I=0\slim (0)}
\left( 1 - \frac{3m^2}{2 (4\pi F)^2}\, \bigl(2g_A^2+1\bigr)
\log\frac{m^2}{\mu^2} \right) 
+ \delta_{k,n-1}^{\phantom{()}}\, E_{T n,n-1}^{I=0\slim (2,\pi)}(t)
+ \ldots \,,
\nonumber \\[0.2em]
E_{T n,k}^{I=1 \phantom{()}} &=
E_{T n,k}^{I=1\slim (0)}
\left( 1 - \frac{m^2}{2 (4\pi F)^2}\,
\left[ \bigl(4g_A^2+1\bigr) \log\frac{m^2}{\mu^2}
  + 4g_A^2 \right] \right) + \ldots \,,
\nonumber \\[0.4em]
\widetilde{E}_{T n,k}^{I=0 \phantom{()}} &=
\widetilde{E}_{T n,k}^{I=0\slim (0)}
\left( 1 - \frac{3m^2}{2 (4\pi F)^2}\, \bigl(2 g_A^2 + 1\bigr)
\log\frac{m^2}{\mu^2} \right) + \ldots \,,
\nonumber \\[0.2em]
\widetilde{E}_{T n,k}^{I=1 \phantom{()}} &=
\widetilde{E}_{T n,k}^{I=1\slim (0)}
\left( 1 - \frac{m^2}{2 (4\pi F)^2}
\left[ \bigl(4g_A^2+1\bigr) \log\frac{m^2}{\mu^2} + 4g_A^2
\right] \right) + \ldots 
\end{align}
and
\begin{align}
  \label{MtI}
\widetilde{M}_{T n,k}^{I=0 \phantom{()}} &=
\widetilde{M}_{T n,k}^{I=0\slim (0)}
\left( 1 - \frac{3m^2}{2 (4\pi F)^2}\,
  \bigl(2g_A^2+1\bigr) \log\frac{m^2}{\mu^2} \right)
+ \widetilde{E}_{T n,k}^{I=0\slim (0)}
  \frac{m^2 g_A^2}{(4\pi F)^2}\, \log\frac{m^2}{\mu^2} 
\nonumber \\[0.4em]
& \quad + \delta_{k,n-2}^{\phantom{I}}\,
          \widetilde{M}_{T n,n-2}^{I=0\slim (2,\pi)}(t)
+ \ldots \,,
\nonumber \\[0.5em]
\widetilde{M}_{T n,k}^{I=1 \phantom{()}} &=
\widetilde{M}_{T n,k}^{I=1\slim (0)} 
\left( 1 - \frac{m^2}{2 (4\pi F)^2}\,
\left[ \left(4g_A^2+1\right) \log\frac{m^2}{\mu^2}
  + 4g_A^2 \right] \right) 
- \widetilde{E}_{n,k}^{I=1\slim (0)}
  \frac{m^2 g_A^2}{3 (4\pi F)^2}\, \log\frac{m^2}{\mu^2}
\nonumber \\[0.4em]
& \quad + \delta_{k,n-1}^{\phantom{I}}\,
          \Big[ \widetilde{M}_{T n,n-1}^{I=1\slim (1,\pi)}(t)
              + \widetilde{M}_{T n,n-1}^{I=1\slim (2,\pi)}(t) \Big]
+ \ldots \,,
\end{align}
where for brevity we have written $\ldots$ to denote analytic terms
proportional to $m^2$ or $t$ and corrections of order $O(q^3)$.  The
analytic terms are due to higher-order tree-level insertions as
specified below \eqref{tree-rules}.  The contributions from pion
operator insertions read
\begin{align}
  \label{Epi}
E_{T n,n-1}^{I=0\slim (2,\pi)}(t) &=
- \frac{1}{4}\slim \widetilde{M}_{T n,n-2}^{I=0\slim (2,\pi)}(t)
\nonumber \\[0.4em]
\widetilde{M}_{T n,n-2}^{I=0\slim (2,\pi)}(t) &=
- \frac{3 g_A^2}{(4\pi F)^2}\,
\sum_{\substack{j=2\\ \mathrm{even}}}^n (j-1)\, M b^{}_{T n,n-j}\, 
\int_{-1}^1 d\eta\, \eta^{j-2}\,
m^2(\eta) \log\frac{m^2(\eta)}{\mu^2}
\end{align}
with $n$ even in the isosinglet case and
\begin{align}
  \label{Mpi}
& M_{T n,n-1}^{I=1\slim (2,\pi)}(t) =
\frac{1}{4 (4\pi F)^2}\,
\sum_{\substack{j=1\\ \mathrm{odd}}}^n M b^{}_{T n,n-j}\, \Bigg\{
  4 g_A^2\slim m^2 \log\frac{m^2}{\mu^2}
\nonumber \\
& \quad + \int_{-1}^1 d\eta\, \eta^{j-1}\, \left[ 
  g_A^2\, \bigl( 2m^2-t \bigr) 
          \left( \log\frac{m^2(\eta)}{\mu^2}+1 \right)
  - (g_A^2 -1)\, m^2(\eta) \log\frac{m^2(\eta)}{\mu^2} \right]
\Bigg\} \,,
\nonumber \\[0.6em]
& \widetilde{M}_{T n,n-1}^{I=1\slim (1,\pi)}(t) =
- \frac{\pi\slim M g_A^2}{(4\pi F)^2}\, 
\sum_{\substack{j=1\\ \mathrm{odd}}}^n M b^{}_{T n,n-j}
\int_{-1}^1d\eta\,\eta^{j-1}\,m(\eta) \,,
\nonumber \\[0.2em]
& \widetilde{M}_{T n,n-1}^{I=1\slim (2,\pi)}(t) =
- \frac{1}{4 (4\pi F)^2}\, 
\sum_{\substack{j=1\\ \mathrm{odd}}}^n M b^{}_{T n,n-j}\, \Bigg\{
  4 g_A^2\slim m^2 \log\frac{m^2}{\mu^2}
\nonumber \\[0.2em]
& \quad + \int_{-1}^1 d\eta\, \eta^{j-1}\, \left[ 
  g_A^2\, \bigl( 2m^2-t \bigr) 
          \left( \log\frac{m^2(\eta)}{\mu^2}+1 \right)
  + \bigl( g_A^2 - 1 - 4M c_4 \bigr)\, 
    m^2(\eta) \log\frac{m^2(\eta)}{\mu^2} \right] \Bigg\}
\end{align}
with $n$ odd in the isotriplet sector.  As remarked in \cite{ando},
the corrections from pion operator insertions are very similar for the
chiral-odd and chiral-even form factors.  We find a correspondence
\begin{align}
  \label{even-odd-corr}
\widetilde{M}_{T n,n-2}^{I=0} & \leftrightarrow
  - M_{n,n-2}^{I=0} \,, &
\widetilde{M}_{T n,n-1}^{I=1} & \leftrightarrow
  \half\slim M_{n,n-1}^{I=1} \,, &
M_{T n,n-1}^{I=1} & \leftrightarrow
  \half\slim E_{n,n-1}^{I=1}
\end{align}
for the terms in \eqref{Epi} and \eqref{Mpi} when interchanging 
$M b^{}_{T n,n-j} \leftrightarrow 2^{-j} j A_{n,n-j}^{\pi\slim (0)}$.

Let us compare our results \eqref{MI} to \eqref{Mpi} to those in
Ref.~\cite{ando}, which gives the corrections of order $O(q)$ for
$\widetilde{M}_{T n,k}^{I=1}$ and of order $O(q^2)$ for all other form
factors.\footnote{%
The tensor form factors in \protect\cite{ando} are related to those
introduced here by $M^T \simeq \widetilde{E}_T$, $E^T \simeq M_T/2$,
$W^T \simeq -4 \widetilde{M}_T$ (all up to terms suppressed by factors
of order $\Delta^2/M^2$) and by $C^T = E_T$.}
We agree with the expressions given there, except for the corrections
from nucleon operator insertions without a factor $g_A^2$ in the
isosinglet form factors, which are absent in \cite{ando}, and for the
corresponding term in $M_{T n,k}^{I=1}$, where we have a different
coefficient.  These corrections are due to the tadpole graph in
Fig.~\ref{fig-1}b, with the pion-nucleon vertex generated by the
two-pion terms in the expansion \eqref{tau-odd} of $\tau^0_{o+}$ and
$\tau^a_{o+}$.  Since this vertex has no spin or momentum structure,
the corresponding corrections must be the same for all form factors
with a given isospin.

\begin{figure}[t]
\psfrag{a}[c][c][0.8]{$P-\frac{\Delta}{2}$}
\psfrag{c}[c][c][0.8]{$P+\frac{\Delta}{2}$}
\psfrag{b}[c][c][0.8]{$l+\frac{\Delta}{2}$}
\psfrag{d}[c][c][0.8]{$l-\frac{\Delta}{2}$}
\psfrag{f}[c][c][0.8]{$l$}
\psfrag{aa}[b][b][1.0]{a} 
\psfrag{bb}[b][b][1.0]{b}
\centerline{\includegraphics[width=10cm]{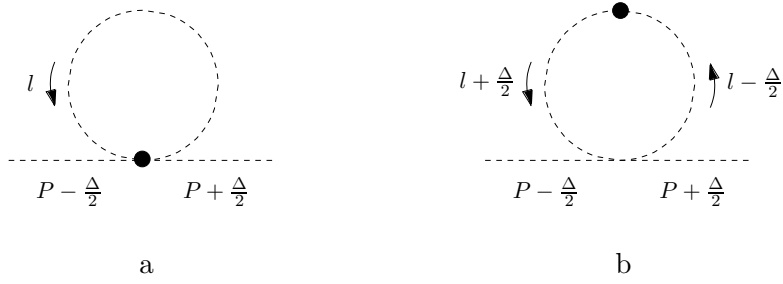}}
\caption{\label{fig-3} One-loop graphs contributing to two-pion matrix
  elements of the pion operator $Q^{A, \lambda\mu}_{n, \pi}(a)$ in
  \protect\eqref{pion-tensor-ops}.  The operator insertion is denoted
  by a black blob.}
\end{figure}

Let us finally give the corrections of order $O(q^2)$ to the
chiral-odd GPDs of the pion.  They are given by the one-loop graphs
shown in Figure~\ref{fig-3} with insertion of the pion operators
\eqref{pion-tensor-ops} and from tree-level insertions of operators
with chiral dimension $n+3$.  For the form factors
\eqref{pion-tensor-ffs} we find
\begin{align}
  \label{pion-tensor-0}
B_{T n,k}^{\pi}(t) &=
B_{T n,k}^{\pi\slim (0)} \left( 1 - \frac{3m^2}{2 (2\pi F)^2}
  \log\frac{m^2}{\mu^2} \right) + \ldots
\\[2mm]
\intertext{for even $n$, and}
  \label{pion-tensor-1}
B_{T n,k}^{\pi}(t) &=
B_{T n,k}^{\pi\slim (0)} \left( 1 - \frac{m^2}{2 (4\pi F)^2}
  \log\frac{m^2}{\mu^2} \right)
\nonumber \\[2mm]
&\quad + \delta_{k,n-1}\, \frac{1}{(4\pi F)^2}\,
\sum_{\substack{j=1\\ \mathrm{odd}}}^n 
     2^{-j}\, B_{T n,n-j}^{\pi\slim (0)}
\int_{-1}^1 d\eta\, \eta^{j-1}\,
     m^2(\eta) \log\frac{m^2(\eta)}{\mu^2} + \ldots
\end{align}	
for odd $n$, where the $\ldots$ stand for analytic terms from
tree-level graphs and for corrections of order $O(q^4)$.  The
corrections going with $\log{m^2/\mu^2}$ are due to the tadpole graph
in Fig.~\ref{fig-3}a and are independent of $k$.  The term involving
an integral over $\eta$ is due to the graph of Fig.~\ref{fig-3}b and
can only occur for $k=n-1$ (and thus only in the isotriplet case).
This is because the operator insertion on the pion line cannot produce
any factor $(aP)$ and the four-pion vertex can only produce one such
factor after the loop integration.

%%%%%%%%%%%%%%%%%%%%%%%%%%%%%%%%%%%%%%%%%%%%%%%%%
%%%%%%%%%%%%%%%%%%%%%%%%%%%%%%%%%%%%%%%%%%%%%%%%%

%%%%%%%%%%%%%%%%%%%%%%%%%%%%%%%%%%%%%%
\section{Results for moments of nucleon and pion GPDs}
\label{sect:gpd-results}
%%%%%%%%%%%%%%%%%%%%%%%%%%%%%%%%%%%%%%

In this section, we rewrite our results in terms of the form factors
$A_{n,k}$, $B_{n,k}$, $C_n$, $\widetilde{A}_{n,k}$,
$\widetilde{B}_{n,k}$ and $A_{T n,k}$, $\widebar{B}_{T n,k}$,
$\widetilde{A}_{T n,k}$, $\widetilde{B}_{T n,k}$, which describe the
moments of GPDs in commonly used parameterizations.  We give
expressions for the value and the first derivative of each form factor
at $t=0$, which should be useful for applications in lattice QCD.  The
corrections obtained from graphs with pion-nucleon operator insertions
are completely specified in this way, because they are independent of
$t$.  In the following we will use the abbreviation $\Lambda_\chi =
4\pi F$.

For a convenient overview of results we also reproduce the expressions
for isosinglet distributions from \cite{DMS2} here.  Together with
\eqref{ff-trafo}, \eqref{ff-inverse} and the expressions in
Section~\ref{sect:isotriplet-results}, we find that up to corrections
of order $O(m^3)$ the chiral-even vector form factors at $t=0$ have
the form
\begin{align}
  \label{res-gpd-vec}
A_{n,k}^{I=0}(0) &= A_{n,k}^{I=0\slim (0)} 
   + A_{n,k}^{I=0\slim (2,m)} m^2 ,
   \phantom{\frac{m^2}{\Lambda_\chi^2}}
\nonumber \\[0.1em]
B_{n,k}^{I=0}(0) &= B_{n,k}^{I=0\slim (0)} 
   - \Bigl( A_{n,k}^{I=0\slim (0)}
     + B_{n,k}^{I=0\slim (0)} \slim\Bigr)\, 
       \frac{3m^2g_A^2}{\Lambda_\chi^2}\, \log\frac{m^2}{\mu^2} 
   + B_{n,k}^{I=0\slim (2,m)} m^2
   + \delta_{k,n-2}^{\phantom{0}}\, B_{n,n-2}^{I=0\slim (2,\pi)}(0) \,,
\nonumber \\[0.1em]
C_{n}^{I=0}(0) &= C_{n}^{I=0\slim (0)} + C_{n}^{I=0\slim (2,m)} m^2
   + C_{n}^{I=0\slim (1,\pi)}(0) + C_{n}^{I=0\slim (2,\pi)}(0) \,,
   \phantom{\frac{m^2}{\Lambda_\chi^2}}
\nonumber \\[0.1em]
A_{n,k}^{I=1}(0) &= A_{n,k}^{I=1\slim (0)}
    \left\{ 1-\frac{m^2}{\Lambda_\chi^2}
    \left[(3g_A^2+1) \log\frac{m^2}{\mu^2} + 2g_A^2\right] \right\}
  + A_{n,k}^{I=1\slim (2,m)}\slim m^2
  + \delta_{k,n-1}^{\phantom{0}}\, A_{n,n-1}^{I=1\slim (2,\pi)}(0) \,,
\nonumber \\[0.1em]
B_{n,k}^{I=1}(0) &= B_{n,k}^{I=1\slim (0)}
    \left\{ 1-\frac{m^2}{\Lambda_\chi^2}
    \left[(3g_A^2+1) \log\frac{m^2}{\mu^2} + 2g_A^2\right] \right\}
  + \Bigl( A_{n,k}^{I=1\slim (0)} 
    + B_{n,k}^{I=1\slim (0)} \slim\Bigr)\,
      \frac{m^2 g_A^2}{\Lambda_\chi^2}\, \log\frac{m^2}{\mu^2}
\nonumber \\[0.1em]
 &\quad + B_{n,k}^{I=1\slim (2,m)}\slim m^2
  + \delta_{k,n-1}^{\phantom{0}}
    \Big[ B_{n,n-1}^{I=1\slim (1,\pi)}(0)
        + B_{n,n-1}^{I=1\slim (2,\pi)}(0) \Big] \,,
  \phantom{\frac{m^2}{\Lambda_\chi^2}}
\nonumber \\[0.1em]
C_{n}^{I=1}(0) &= C_{n}^{I=1\slim (0)}
    \left\{ 1-\frac{m^2}{\Lambda_\chi^2}
    \left[(3g_A^2+1) \log\frac{m^2}{\mu^2} + 2g_A^2\right] \right\}
  + C_{n}^{I=1\slim (2,m)}\slim m^2 .
\end{align}
The labeling of coefficients with superscripts $(0)$, $(2,m)$,
$(1,\pi)$ and $(2,\pi)$ follows the same pattern as in
Sections~\ref{sect:isotriplet-results} and \ref{sect:tensor-results}.
The contributions from pion operator insertions read
\begin{align}
  \label{gpd-vec0-pi}
B_{n,n-2}^{I=0\slim (2,\pi)}(0) &=
\frac{6m^2 g_A^2}{\Lambda_\chi^2}\,
  \log\frac{m^2}{\mu^2}\,
  \sum_{\substack{j=2\\ \mathrm{even}}}^{n} 
  2^{-j} j\, A_{n,n-j}^{\pi\slim (0)} \,,
\nonumber \\[0.1em]
C_{n}^{I=0\slim (1,\pi)}(0) &= 
\frac{3\pi\slim m M g_A^2}{2 \Lambda_\chi^2}\,
  \sum_{\substack{j=2\\ \mathrm{even}}}^{n}
  2^{-j}\, \frac{j}{j+1}\, A_{n,n-j}^{\pi\slim (0)} \,,
\nonumber \\[0.1em]
C_{n}^{I=0\slim (2,\pi)}(0) &= 
{}- \frac{3m^2 g_A^2}{2 \Lambda_\chi^2}\,
  \log\frac{m^2}{\mu^2}\,
  \sum_{\substack{j=2\\ \mathrm{even}}}^{n} 
  2^{-j} j\, A_{n,n-j}^{\pi\slim (0)}
+ \frac{12 m^2}{\Lambda_\chi^2}\,
    \Biggl[\, \frac{g_A^2}{8}
    - M c_1 \left(\log\frac{m^2}{\mu^2}+1\right)
\nonumber \\[0.1em]
& \qquad\qquad\quad
    + \frac{3}{4}\slim M c_2\slim \log\frac{m^2}{\mu^2}
    + M c_3 \left(\log\frac{m^2}{\mu^2}+\frac{1}{2}\right)
    \Biggr]\,
    \sum_{\substack{j=2\\ \mathrm{even}}}^{n} 
    2^{-j} \frac{j}{j+1}\, A_{n,n-j}^{\pi\slim (0)} \,,
\\
\intertext{where $n$ is even, and}
  \label{gpd-vec1-pi}
A_{n,n-1}^{I=1\slim (2,\pi)}(0) &=
  \frac{2 m^2 g_A^2}{\Lambda_\chi^2}\, \log\frac{m^2}{\mu^2}\,
  \sum_{\substack{j=1\\ \mathrm{odd}}}^{n}
    2^{-j} j\, A_{n,n-j}^{\pi\slim (0)}
+ \frac{m^2}{\Lambda_\chi^2}
     \left[ (g_A^2+1) \log\frac{m^2}{\mu^2} + 2 g_A^2 \right]
  \sum_{\substack{j=1\\ \mathrm{odd}}}^{n} 
    2^{-j} A_{n,n-j}^{\pi\slim (0)} \,,
\nonumber \\[0.1em]
B_{n,n-1}^{I=1\slim (1,\pi)}(0) &=
- \frac{4\pi\slim m M g_A^2}{\Lambda_\chi^2}\,
  \sum_{\substack{j=1\\ \mathrm{odd}}}^n
  2^{-j} A_{n,n-j}^{\pi\slim (0)} \,,
\nonumber \\[0.1em]
B_{n,n-1}^{I=1\slim (2,\pi)}(0) &=
- \frac{4 m^2 g_A^2}{\Lambda_\chi^2}\, \log\frac{m^2}{\mu^2}\,
  \sum_{\substack{j=1\\ \mathrm{odd}}}^{n}
     2^{-j} j\, A_{n,n-j}^{\pi\slim (0)}
\nonumber \\
& \quad - \frac{4 m^2}{\Lambda_\chi^2}
     \Biggl[\, g_A^2 \left( \log\frac{m^2}{\mu^2} + 1 \right)
          - M c_4 \log\frac{m^2}{\mu^2} \Biggr]\,
  \sum_{\substack{j=1\\ \mathrm{odd}}}^{n} 
     2^{-j} A_{n,n-j}^{\pi\slim (0)} \,,
\end{align}
where $n$ is odd.  The $A_{n,n-j}^{\pi\slim (0)}$ are the leading
terms in the chiral expansion of the pion form factors in
\eqref{pion-gpds} and fulfill the relations \cite{DMS1}
\begin{align}
  \label{soft-pion-th}
\sum_{\substack{j=2\\ \mathrm{even}}}^n
  2^{-j} A_{n,n-j}^{\pi\slim (0)} &= 
    - A_{n,n\phantom{j}}^{\pi\slim (0)}
\quad \text{for even $n$,}
&
\sum_{\substack{j=1\\ \mathrm{odd}}}^n
  2^{-j} A_{n,n-j}^{\pi\slim (0)} &= B_{n\phantom{j}}^{\pi\slim (0)}
\quad \text{for odd $n$,}
\end{align}
where $B_n^{\pi (0)}$ is the $n$-th moment of the pion distribution
amplitude to leading order in the chiral expansion, as introduced in
Section~\ref{sect:axial}.  Estimates for the values of the low-energy
constants $c_i$ appearing in \eqref{gpd-vec0-pi} and
\eqref{gpd-vec1-pi} can be found in \cite{Meissner:2005ba}.

For the axial form factors we have
\begin{align}
  \label{res-gpd-ax}
\widetilde{A}_{n,k}^{I=0}(0) &= \widetilde{A}_{n,k}^{I=0\slim (0)}\, 
  \left\{ 1-\frac{3m^2g_A^2}{\Lambda_\chi^2}
  \left[\log\frac{m^2}{\mu^2}+1\right] \right\}
  + \widetilde{A}_{n,k}^{I=0\slim (2,m)} m^2 ,
\nonumber \\[0.1em]
\widetilde{B}_{n,k}^{I=0}(0) &= \widetilde{B}_{n,k}^{I=0\slim (0)}\, 
  \left\{ 1-\frac{3m^2g_A^2}{\Lambda_\chi^2}
  \left[\log\frac{m^2}{\mu^2}+1\right] \right\}
  - \widetilde{A}_{n,k}^{I=0\slim (0)}\; \frac{m^2g_A^2}{\Lambda_\chi^2}\,
    \log\frac{m^2}{\mu^2}
  + \widetilde{B}_{n,k}^{I=0\slim (2,m)} m^2 ,
\displaybreak
\nonumber \\[0.1em]
\widetilde{A}_{n,k}^{I=1}(0)
 &= \widetilde{A}_{n,k}^{I=1\slim (0)}
    \left\{ 1-\frac{m^2}{\Lambda_\chi^2}
    \left[(2g_A^2+1) \log\frac{m^2}{\mu^2} + {g_A^2}\right] \right\}
  + \widetilde{A}_{n,k}^{I=1\slim (2,m)}\slim m^2 ,
\nonumber \\[0.1em]
\widetilde{B}_{n,k}^{I=1}(0) &= \widetilde{B}_{n,k}^{I=1\slim (0)}
    \left\{ 1-\frac{m^2}{\Lambda_\chi^2}
    \left[(2g_A^2+1) \log\frac{m^2}{\mu^2} + {g_A^2}\right] \right\}
\nonumber \\[0.2em]
 & \quad
  + \widetilde{A}_{n,k}^{I=1\slim (0)}\, 
    \frac{m^2 g_A^2}{3 \Lambda_\chi^2}\, \log\frac{m^2}{\mu^2}
  + \widetilde{B}_{n,k}^{I=1\slim (2,m)}\slim m^2 
\hspace{8em} \text{for $k< n-1$}
\\
\intertext{with corrections of order $O(m^3)$, and}
  \label{Bti0}
\widetilde{B}_{n,n-1}^{I=1}(0) &= 
B_n^\pi\, \frac{4 M^2 g_A}{m_\pi^2} \,
  \bigl( 1 - 2m_\pi^2\slim g_A^{-1} d_{18}^{} \bigr)
+ \widetilde{B}_{n,n-1}^{I=1\slim (0)} + O(m) 
\hspace{4.4em} \text{for odd $n$.}
\end{align}
Note that the implicit pion mass dependence from $B_n^\pi$, $M$, $g_A$
and $m_\pi$ is relevant within the accuracy of this expression.
Numerical estimates of the low-energy constant $d_{18}$ are given in
\cite{Fettes:2000xg}.
The derivative of $\widetilde{B}_{n,n-1}^{I=1}(t)$ at $t=0$ reads
\begin{align}
\partial_t^{} \widetilde{B}_{n,n-1}^{I=1}(0) &= 
B_n^\pi\, \frac{4 M^2 g_A}{m_\pi^4} \,
  \bigl( 1 - 2m_\pi^2\slim g_A^{-1} d_{18}^{} \bigr) + O(m^{-1}) \,,
\end{align}
where the order $O(m^{-1})$ corrections are due to terms of the form
$O(q^3) /(m_\pi^2 -t)$ in $\widetilde{B}_{n,n-1}^{I=1}(t)$.  Using
$B_1^\pi=1$, we obtain a ratio
\begin{equation}
\frac{\partial_t^{} \widetilde{B}_{n,n-1}^{I=1}(0)}{%
      \partial_t^{} \widetilde{B}_{1,0}^{I=1}(0) \rule{0pt}{1.1em}}
= B_n^\pi + O(m^3)
\end{equation}
which involves only physical matrix elements and is independent of any
low-energy constants.  It would be interesting to test this relation
in lattice QCD calculations, as this would indicate how well the
chiral expansion works at a given pion mass.

The derivatives at $t=0$ of the remaining chiral-even form factors
have nonanalytic contributions in the pion mass only for
\begin{align}
\partial_t^{} B_{n,n-2}^{I=0}(0) &= B_{n,n-2}^{I=0\slim (2,t)} 
  + \partial_t^{} B_{n,n-2}^{I=0\slim (2,\pi)}(0) \,,
\nonumber \\[0.5em]
\partial_t^{} C_{n}^{I=0}(0) &= C_{n}^{I=0\slim (2,t)}
  + \partial_t^{} C_{n}^{I=0\slim (1,\pi)}(0) 
  + \partial_t^{} C_{n}^{I=0\slim (2,\pi)}(0)
\nonumber \\[0.3em]
\partial_t^{} A_{n,n-1}^{I=1}(0) &= A_{n,n-1}^{I=1\slim (2,t)} 
  + \partial_t^{} A_{n,n-1}^{I=1\slim (2,\pi)}(0) \,,
\nonumber \\[0.3em]
\partial_t^{} B_{n,n-1}^{I=1}(0) &= B_{n,n-1}^{I=1\slim (2,t)}
  + \partial_t^{} B_{n,n-1}^{I=1\slim (1,\pi)}(0) 
  + \partial_t^{} B_{n,n-1}^{I=1\slim (2,\pi)}(0) \,,
\end{align}
with corrections of order $O(m)$, where
\begin{align}
  \label{gpd-vec0-pi-t}
\partial_t^{} B_{n,n-2}^{I=0\slim (2,\pi)}(0) &= 
{}-\frac{3g_A^2}{\Lambda_\chi^2}
  \left(\log\frac{m^2}{\mu^2}+1\right)\,
  \sum_{\substack{j=2\\ \mathrm{even}}}^{n}
  2^{-j} \frac{j}{j+1}\, A_{n,n-j}^{\pi\slim (0)} \,,
\nonumber \\[0.1em]
\partial_t^{} C_{n}^{I=0\slim (1,\pi)}(0) &= 
{}- \frac{M}{m}\, \frac{\pi g_A^2}{8 \Lambda_\chi^2}\,
  \sum_{\substack{j=2\\ \mathrm{even}}}^{n}
  2^{-j}\, \frac{j\, (5j+14)}{(j+1) (j+3)}\,
  A_{n,n-j}^{\pi\slim (0)} \,,
\displaybreak
\nonumber \\[0.1em]
\partial_t^{} C_{n}^{I=0\slim (2,\pi)}(0) &= 
{}- \frac{3g_A^2}{4 \Lambda_\chi^2}\,
  \left(\log\frac{m^2}{\mu^2} +3\right)\,
  \sum_{\substack{j=2\\ \mathrm{even}}}^{n} 
  2^{-j} \frac{j}{j+1}\, A_{n,n-j}^{\pi\slim (0)}\,
+ \frac{2}{\Lambda_\chi^2}\,
  \Biggl[\, \frac{g_A^2}{8} + M c_1 
\nonumber \\[0.1em]
 & \hspace{-1em}
    - \frac{3}{4}\slim M c_2 \left(\log\frac{m^2}{\mu^2}+1\right)
    - M c_3 \left(\log\frac{m^2}{\mu^2}+\frac{3}{2}\right)
    \Biggr]\,
  \sum_{\substack{j=2\\ \mathrm{even}}}^{n} 
  2^{-j} \frac{j\, (j+4)}{(j+1) (j+3)}\, A_{n,n-j}^{\pi\slim (0)}
\\
\intertext{with $n$ even, and}
  \label{gpd-vec1-pi-t}
\partial_t^{} A_{n,n-1}^{I=1\slim (2,\pi)}(0) &= 
- \frac{g_A^2}{\Lambda_\chi^2}
  \left(\log\frac{m^2}{\mu^2}+1\right)\,
  \sum_{\substack{j=1\\ \mathrm{odd}}}^{n}
  2^{-j} A_{n,n-j}^{\pi\slim (0)}
\nonumber \\
& \quad + \frac{1}{2 \Lambda_\chi^2}
  \left[ (g_A^2-1)\log\frac{m^2}{\mu^2} - (g_A^2+1) \right]\,
  \sum_{\substack{j=1\\ \mathrm{odd}}}^{n}
  2^{-j}\, \frac{1}{j+2}\, A_{n,n-j}^{\pi\slim (0)} \,,
\nonumber \\[0.2em]
\partial_t^{} B_{n,n-1}^{I=1\slim (1,\pi)}(0) &= 
  \frac{M}{m}\, \frac{\pi g_A^2}{\Lambda_\chi^2}\,
  \sum^n_{\substack{j=1\\ \mathrm{odd}}}
  2^{-j}\, \frac{1}{j+2}\, A^{\pi\slim (0)}_{n,n-j} \,,
\nonumber \\[0.4em]
\partial_t^{} B_{n,n-1}^{I=1\slim (2,\pi)}(0) &= 
  \frac{2 g_A^2}{\Lambda_\chi^2}
  \left(\log\frac{m^2}{\mu^2}+1\right)\,
  \sum_{\substack{j=1\\ \mathrm{odd}}}^{n} 
  2^{-j} A_{n,n-j}^{\pi\slim (0)}
\nonumber \\
& \quad + \frac{2}{\Lambda_\chi^2}
  \Biggl[\, g_A^2 - M c_4 \left(\log\frac{m^2}{\mu^2}+1\right) \Biggr]\,
  \sum_{\substack{j=1\\ \mathrm{odd}}}^{n} 
  2^{-j}\frac{1}{j+2}\, A_{n,n-j}^{\pi\slim (0)} \,.
\end{align}
with $n$ odd.  All other chiral-even nucleon form factors receive only
corrections from pion-nucleon operators, so that their derivatives at
$t=0$ are given by the appropriate coefficients with superscript
$(2,t)$, which are due to tree-level contributions.

For the chiral-odd nucleon form factors at $t=0$ we find
\begin{align}
  \label{res-gpd-ten}
A_{T n,k}^{I=0 \phantom{()}}(0) &=
A_{T n,k}^{I=0\slim (0)}
\left\{ 1 - \frac{3m^2}{2 \Lambda_\chi^2}\, \bigl(2 g_A^2 + 1\bigr)
  \log\frac{m^2}{\mu^2} \right\} 
+ A_{T n,k}^{I=0\slim (2,m)} m^2 ,
\nonumber \\[0.2em]
\widebar{B}_{T n,k}^{I=0 \phantom{()}}(0) &=
\widebar{B}_{T n,k}^{I=0\slim (0)}
\left\{ 1 - \frac{3m^2}{2 \Lambda_\chi^2}\, \bigl(2 g_A^2 + 1\bigr)
  \log\frac{m^2}{\mu^2} \right\}
\nonumber \\[0.2em]
& \quad 
+ \Bigl( A_{T n,k}^{I=0\slim (0)}
       + \widebar{B}_{T n,k}^{I=0\slim (0)} \slim\Bigr)\,
    \frac{3 m^2 g_A^2}{\Lambda_\chi^2}\, \log\frac{m^2}{\mu^2}
+ \widebar{B}_{T n,k}^{I=0\slim (2,m)} m^2 ,
  \phantom{\frac{m^2}{\Lambda_\chi^2}}
\nonumber \\[0.2em]
\widetilde{A}_{T n,k}^{I=0 \phantom{()}}(0) &=
\widetilde{A}_{T n,k}^{I=0\slim (0)}
\left\{ 1 - \frac{3 m^2}{2 \Lambda_\chi^2}\,
       \bigl(2g_A^2+1\bigr) \log\frac{m^2}{\mu^2} \right\}
  + \left( A_{T n,k}^{I=0\slim (0)}
         + \frac{3}{2}\slim \widebar{B}_{T n,k}^{I=0\slim (0)} \right)
    \frac{m^2 g_A^2}{\Lambda_\chi^2}\, \log\frac{m^2}{\mu^2}
\nonumber \\[0.2em]
& \quad 
  + \widetilde{A}_{T n,k}^{I=0\slim (2,m)} m^2 
  + \delta_{k,n-2}^{\phantom{I}}\,
          \widetilde{A}_{T n,n-2}^{I=0\slim (2,\pi)}(0) \,,
  \phantom{\frac{m^2}{\Lambda_\chi^2}}
\nonumber \\[0.2em]
\widetilde{B}_{T n,k}^{I=0 \phantom{()}}(0) &=
\widetilde{B}_{T n,k}^{I=0\slim (0)}
\left\{ 1 - \frac{3m^2}{2 \Lambda_\chi^2}\, \bigl(2g_A^2+1\bigr)
\log\frac{m^2}{\mu^2} \right\}
  + \widetilde{B}_{T n,k}^{I=0\slim (2,m)} m^2
  + \delta_{k,n-1}^{\phantom{I}}\,
       \widetilde{B}_{T n,n-1}^{I=0\slim (2,\pi)}(0) \,,
\displaybreak
\nonumber \\[0.2em]
A_{T n,k}^{I=1 \phantom{()}}(0) &=
A_{T n,k}^{I=1\slim (0)}
\left\{ 1 - \frac{m^2}{2 \Lambda_\chi^2}
  \left[ \bigl(4g_A^2+1\bigr) \log\frac{m^2}{\mu^2} + 4g_A^2
  \right] \right\}
+ A_{T n,k}^{I=1\slim (2,m)} m^2 ,
\nonumber \\[0.2em]
\widebar{B}_{T n,k}^{I=1 \phantom{()}}(0) &=
\widebar{B}_{T n,k}^{I=1\slim (0)}
\left\{ 1 - \frac{m^2}{2 \Lambda_\chi^2}\,
  \left[ \bigl(4g_A^2+1\bigr) \log\frac{m^2}{\mu^2} + 4 g_A^2
  \right] \right\} 
\nonumber \\[0.2em]
& \quad 
- \Bigl( A_{T n,k}^{I=1\slim (0)} 
       + \widebar{B}_{T n,k}^{I=1\slim (0)} \slim\Bigr)\,
    \frac{m^2 g_A^2}{\Lambda_\chi^2}\, \log\frac{m^2}{\mu^2}
  + \widebar{B}_{T n,k}^{I=1\slim (2,m)} m^2 
  + \delta_{k,n-1}^{\phantom{I}}\,
    \widebar{B}_{T n,n-1}^{I=1\slim (2,\pi)}(0) \,,
  \phantom{\frac{m^2}{\Lambda_\chi^2}}
\nonumber \\[0.2em]
\widetilde{A}_{T n,k}^{I=1 \phantom{()}}(0) &=
\widetilde{A}_{T n,k}^{I=1\slim (0)}
\left\{ 1 - \frac{m^2}{2 \Lambda_\chi^2}
\left[ \bigl(4g_A^2+1\bigr) \log\frac{m^2}{\mu^2} + 4g_A^2 \right]
\right\}
  - \left( A_{T n,k}^{I=1\slim (0)}
         + \frac{3}{2}\slim \widebar{B}_{T n,k}^{I=1\slim (0)} \right)
    \frac{m^2 g_A^2}{3 \Lambda_\chi^2}\, \log\frac{m^2}{\mu^2} 
\nonumber \\[0.2em]
& \quad
  + \widetilde{A}_{T n,k}^{I=1\slim (2,m)} m^2 
  + \delta_{k,n-1}^{\phantom{I}} 
          \Big[ \widetilde{A}_{T n,n-1}^{I=1\slim (1,\pi)}(0)
              + \widetilde{A}_{T n,n-1}^{I=1\slim (2,\pi)}(0) \Big] \,,
  \phantom{\frac{m^2}{\Lambda_\chi^2}}
\nonumber \\[0.2em]
\widetilde{B}_{T n,k}^{I=1 \phantom{()}}(0) &=
\widetilde{B}_{T n,k}^{I=1\slim (0)}
\left\{ 1 - \frac{m^2}{2 \Lambda_\chi^2}\,
\left[ \bigl(4g_A^2+1\bigr) \log\frac{m^2}{\mu^2}
  + 4g_A^2 \right] \right\}
  + \widetilde{B}_{T n,k}^{I=1\slim (2,m)} m^2 
\end{align}
with corrections of order $O(m^3)$, where
\begin{align}
  \label{gpd-ten0-pi}
\widetilde{A}_{T n,n-2}^{I=0\slim (2,\pi)}(0) &=
  \frac{3m^2 g_A^2}{\Lambda_\chi^2}\, \log\frac{m^2}{\mu^2} 
  \sum_{\substack{j=2\\ \mathrm{even}}}^{n} M b^{}_{T n,n-j} \,,
\nonumber \\
\widetilde{B}_{T n,n-1}^{I=0\slim (2,\pi)}(0) &=
  \frac{1}{2}\slim \widetilde{A}_{T n,n-2}^{I=0\slim (2,\pi)}(0)
\\
\intertext{with $n$ even, and}
  \label{gpd-ten1-pi}
\widebar{B}_{T n,n-1}^{I=1\slim (2,\pi)}(0) &=
  \frac{m^2 g_A^2}{\Lambda_\chi^2}\, \log\frac{m^2}{\mu^2}
  \sum_{\substack{j=1\\ \mathrm{odd}}}^{n} M b^{}_{T n,n-j}
+ \frac{m^2}{2 \Lambda_\chi^2}\, 
  \left[ \bigl(g_A^2+1\bigr) \log\frac{m^2}{\mu^2} + 2g_A^2 \right]
  \sum_{\substack{j=1\\ \mathrm{odd}}}^{n} \frac{1}{j}\, 
    M b^{}_{T n,n-j} \,,
\nonumber \\
\widetilde{A}_{T n,n-1}^{I=1\slim (1,\pi)}(0) &=
  \frac{\pi\slim m M g_A^2}{\Lambda_\chi^2}\,
  \sum_{\substack{j=1\\ \mathrm{odd}}}^{n}\, \frac{1}{j}\, 
    M b^{}_{T n,n-j} \,,
\nonumber \\
\widetilde{A}_{T n,n-1}^{I=1\slim (2,\pi)}(0) &=
  \frac{m^2 g_A^2}{\Lambda_\chi^2}\, \log\frac{m^2}{\mu^2}
  \sum_{\substack{j=1\\ \mathrm{odd}}}^{n} M b^{}_{T n,n-j}
\nonumber \\
& \quad + \frac{m^2}{\Lambda_\chi^2}
      \Biggl[\, g_A^2 \left( \log\frac{m^2}{\mu^2} + 1 \right)
          - M c_4 \log\frac{m^2}{\mu^2} \Biggr]\,
  \sum_{\substack{j=1\\ \mathrm{odd}}}^{n} \frac{1}{j}\, 
    M b^{}_{T n,n-j}
\end{align}
with $n$ odd.  Our results for $A_{n,0}^{I=1}(0)$,
$\widetilde{A}_{n,0}^{I=1}(0)$ and $A_{T n,0}^{I=1}(0)$ reproduce the
expressions in \cite{Chen:2001eg} for the distributions of
unpolarized, longitudinally and transversely polarized quarks and
antiquarks in the nucleon.
The derivatives at $t=0$ of the following form factors have
nonanalytic contributions in the pion mass:
\begin{align}
\partial_t^{} \widetilde{A}_{T n,n-2}^{I=0}(0) &= 
  \widetilde{A}_{T n,n-2}^{I=0\slim (2,t)} 
  + \partial_t^{} \widetilde{A}_{T n,n-2}^{I=0\slim (2,\pi)}(0) \,,
\nonumber \\[0.3em]
\partial_t^{} \widetilde{B}_{T n,n-1}^{I=0\phantom{()}}(0) &= 
    \widetilde{B}_{T n,n-1}^{I=0\slim (2,t)}
  + \partial_t^{} \widetilde{B}_{T n,n-1}^{I=0\slim (2,\pi)}(0) \,,
\nonumber \\[0.3em]
\partial_t^{} \widebar{B}_{T n,n-1}^{I=1}(0) &= 
  \widebar{B}_{T n,n-1}^{I=1\slim (2,t)} 
  + \partial_t^{} \widebar{B}_{T n,n-1}^{I=1\slim (2,\pi)}(0) \,,
\nonumber \\[0.3em]
\partial_t^{} \widetilde{A}_{T n,n-1}^{I=1\phantom{()}}(0) &= 
    \widetilde{A}_{T n,n-1}^{I=1\slim (2,t)}
  + \partial_t^{} \widetilde{A}_{T n,n-1}^{I=1\slim (1,\pi)}(0)
  + \partial_t^{} \widetilde{A}_{T n,n-1}^{I=1\slim (2,\pi)}(0) \,,
\end{align}
where corrections are of order $O(m)$ and
\begin{align}
  \label{gpd-ten0-pi-t}
\partial_t^{} \widetilde{A}_{T n,n-2}^{I=0\slim (2,\pi)}(0) &=
  - \frac{3 g_A^2}{2 \Lambda_\chi^2}
    \left( \log\frac{m^2}{\mu^2} + 1 \right)
    \sum_{\substack{j=2\\ \mathrm{even}}}^{n} 
       \frac{1}{j+1}\, M b^{}_{T n,n-j} \,,
\nonumber \\
\partial_t^{} \widetilde{B}_{T n,n-1}^{I=0\slim (2,\pi)}(0) &=
  \frac{1}{2}\slim
     \partial_t^{} \widetilde{A}_{T n,n-2}^{I=0\slim (2,\pi)}(0)
\\
\intertext{with even $n$ and}
  \label{gpd-ten1-pi-t}
\partial_t^{} \widebar{B}_{T n,n-1}^{I=1\slim (2,\pi)}(0) &=
  - \frac{g_A^2}{2 \Lambda_\chi^2}
    \left( \log\frac{m^2}{\mu^2} + 1 \right)
    \sum_{\substack{j=1\\ \mathrm{odd}}}^{n} 
       \frac{1}{j}\, M b^{}_{T n,n-j}
\nonumber \\[0.2em]
& \quad + \frac{1}{4 \Lambda_\chi^2}
    \left[ (g_A^2-1) \log\frac{m^2}{\mu^2} - (g_A^2+1) \right]
    \sum_{\substack{j=1\\ \mathrm{odd}}}^{n} 
       \frac{1}{j (j+2)}\, M b^{}_{T n,n-j} \,,
\nonumber \\
\partial_t^{} \widetilde{A}_{T n,n-1}^{I=1\slim (1,\pi)}(0) &=
  - \frac{M}{m}\, \frac{\pi g_A^2}{4 \Lambda_\chi^2}\,
  \sum^n_{\substack{j=1\\ \mathrm{odd}}}
    \frac{1}{j(j+2)}\, M b^{}_{T n,n-j} \,,
\nonumber \\
\partial_t^{} \widetilde{A}_{T n,n-1}^{I=1\slim (2,\pi)}(0) &=
  - \frac{g_A^2}{2 \Lambda_\chi^2}
    \left( \log\frac{m^2}{\mu^2} + 1 \right)
    \sum^n_{\substack{j=1\\ \mathrm{odd}}} \frac{1}{j}\, 
      M b^{}_{T n,n-j}
\nonumber \\[0.2em]
& \quad - \frac{1}{2 \Lambda_\chi^2}
  \Biggl[\, g_A^2 - M c_4 \left(\log\frac{m^2}{\mu^2}+1\right) \Biggr]\,
  \sum^n_{\substack{j=1\\ \mathrm{odd}}}
       \frac{1}{j (j+2)}\, M b^{}_{T n,n-j}
\end{align}
with odd $n$.  As a consequence of the relations
\eqref{even-odd-corr}, we find the following correspondence between
the corrections \eqref{gpd-ten0-pi}, \eqref{gpd-ten1-pi},
\eqref{gpd-ten0-pi-t}, \eqref{gpd-ten1-pi-t} from pion loop insertions
to chiral-odd form factors and their chiral-even counterparts
\eqref{gpd-vec0-pi}, \eqref{gpd-vec1-pi}, \eqref{gpd-vec0-pi-t},
\eqref{gpd-vec1-pi-t}:
\begin{align}
\widetilde{A}_{T n,n-2}^{I=0} & \leftrightarrow
  \half\slim B_{n,n-2}^{I=0} &
\widetilde{A}_{T n,n-1}^{I=1} & \leftrightarrow
  - \tfrac{1}{4}\slim B_{n,n-1}^{I=1} &
\widebar{B}\slim{}_{T n,n-1}^{I=1} & \leftrightarrow
  \half\slim A_{n,n-1}^{I=1}
\end{align}
when the low-energy constants are interchanged as $M b^{}_{T n,n-j}
\leftrightarrow 2^{-j} j A_{n,n-j}^{\pi\slim (0)}$.

Let us also give the expressions of form factors and their derivatives
at $t=0$ for the moments of pion GPDs.  For the chiral-even moments,
the expressions given in \cite{DMS1} result in
\begin{equation}
  \label{pi-gpd-even}
A_{n,k}^{\pi}(0) =
\begin{cases}
  A_{n,k}^{\pi\slim (0)}
   + A_{n,k}^{\pi\slim (2,m)} m^2 
   + \delta_{k,n}^{\phantom{0}}\, A_{n,n\phantom{1}}^{\pi\slim (l,2)}
  & \quad \text{for even $n$,}
\\[0.6em]
  A_{n,k}^{\pi\slim (0)} \left( 
     1 - \dfrac{m^2}{\Lambda_\chi^2} \log\dfrac{m^2}{\mu^2} \right)
   + A_{n,k}^{\pi\slim (2,m)} m^2
   + \delta_{k,n-1}^{\phantom{0}}\, A_{n,n-1}^{\pi\slim (l,2)} 
  & \quad \text{for odd $n$,}
\end{cases}
\end{equation}
with corrections of order $O(m^4)$ and
\begin{align}
  \label{pi-gpd-pi}
A_{n,n\phantom{1}}^{\pi\slim (l,2)} &= 
- \frac{m^2}{2 \Lambda_\chi^2}
  \left( \log\frac{m^2}{\mu^2} + 1 \right)
  \sum_{\substack{j=2\\ \mathrm{even}}}^{n}
    2^{-j} \frac{j}{j+1}\, A_{n,n-j}^{\pi\slim (0)} \,,
\nonumber \\
A_{n,n-1}^{\pi\slim (l,2)} &= \frac{2 m^2}{\Lambda_\chi^2}
  \log\frac{m^2}{\mu^2}
  \sum_{\substack{j=1\\ \mathrm{odd}}}^{n} 
     2^{-j} A_{n,n-j}^{\pi\slim (0)} \,.
\end{align}
Using the relation \eqref{soft-pion-th} one thus has
\begin{equation}
A_{n,n-1}^{\pi \phantom{()}}(0) = A_{n,n-1}^{\pi\slim (0)}
  + \frac{m^2}{\Lambda_\chi^2} \log\frac{m^2}{\mu^2}
    \Bigl[ 2 B_{n\phantom{1}}^{\pi\slim (0)} 
         - A_{n,n-1}^{\pi\slim (0)} \Bigr]
  + A_{n,n-1}^{\pi\slim (2,m)} m^2 + O(m^4)
\end{equation}
with $n$ odd.
For the chiral-odd moments we have with \eqref{pion-tensor-0} and
\eqref{pion-tensor-1}
\begin{equation}
B_{T n,k}^{\pi}(0) =
\begin{cases}
  B_{T n,k}^{\pi\slim (0)}\left( 
     1 - \dfrac{3 m^2}{2 \Lambda_\chi^2} \log\dfrac{m^2}{\mu^2} \right)
   + B_{T n,k}^{\pi\slim (2,m)} m^2 
  & \quad \text{for even $n$,}
\\[1.2em]
  B_{T n,k}^{\pi\slim (0)} \left( 
     1 - \dfrac{m^2}{2 \Lambda_\chi^2} \log\dfrac{m^2}{\mu^2} \right)
   + B_{T n,k}^{\pi\slim (2,m)} m^2
   + \delta_{k,n-1}^{\phantom{0}}\, B_{T n,n-1}^{\pi\slim (l,2)} 
  & \quad \text{for odd $n$,}
\end{cases}
\end{equation}
where corrections are again of order $O(m^4)$ and
\begin{align}
B_{T n,n-1}^{\pi\slim (l,2)} &= \frac{2 m^2}{\Lambda_\chi^2}
  \log\frac{m^2}{\mu^2}
  \sum_{\substack{j=1\\ \mathrm{odd}}}^{n} 
     2^{-j} \frac{1}{j}\, B_{T n,n-j}^{\pi\slim (0)} \,.
\end{align}
The only nonanalytic contributions in the pion mass for the
derivatives of form factors are
\begin{align}
\partial_t^{} A_{n,n}^{\pi}(0) &=
  A_{n,n}^{\pi\slim (2,t)}
+ \frac{1}{\Lambda_\chi^2}
  \left( \log\frac{m^2}{\mu^2} + 1 \right)
  \sum_{\substack{j=2\\ \mathrm{even}}}^{n}
    2^{-j} \frac{j}{j+1}\, A_{n,n-j}^{\pi\slim (0)}
+ \frac{1}{12 \Lambda_\chi^2}
  \sum_{\substack{j=2\\ \mathrm{even}}}^{n}
    2^{-j} \frac{j\, (j+4)}{(j+1) (j+3)}\, A_{n,n-j}^{\pi\slim (0)}
\nonumber \\
\partial_t^{} A_{n,n-1}^{\pi \phantom{()}}(0) &=
  A_{n,n-1}^{\pi\slim (2,t)} - \frac{1}{\Lambda_\chi^2}
     \left( \log\frac{m^2}{\mu^2} + 1 \right)
     \sum_{\substack{j=1\\ \mathrm{odd}}}^{n} 
       2^{-j} \frac{1}{j+2}\, A_{n,n-j}^{\pi\slim (0)} \,,
\nonumber \\
\partial_t^{} B_{T n,n-1}^{\pi \phantom{()}}(0) &=
  B_{T n,n-1}^{\pi\slim (2,t)} - \frac{1}{\Lambda_\chi^2}
     \left( \log\frac{m^2}{\mu^2} + 1 \right)
     \sum_{\substack{j=1\\ \mathrm{odd}}}^{n} 
       2^{-j} \frac{1}{j (j+2)}\, B_{T n,n-j}^{\pi\slim (0)} \,,
\end{align}
where the second index is always even and corrections are of order
$O(m^2)$.

Let us finally take a look at moments of parton distributions whose
values are fixed by quantum numbers for arbitrary values of the pion
mass, see e.g.\ \cite{Chen:2001eg,Arndt:2001ye}.  For
$A_{1,0}^{\pi}(0)$ one readily finds that the explicit chiral
logarithm in \eqref{pi-gpd-even} cancels against the one in
\eqref{pi-gpd-pi}.  This is required to ensure the quark number sum
rule
\begin{equation}
A_{1,0}^{\pi}(0) = \int_0^1 dx\, \bigl( 
   u_\pi - \bar{u}_\pi - d_\pi + \bar{d}_\pi \bigr) = 2 
\end{equation}
in the pion, which also implies $A_{1,0}^{\pi\slim (0)} = 2$ and
$A_{1,0}^{\pi\slim (2,m)} = 0$.  With this one also finds that our
result \eqref{res-gpd-vec} is consistent with the quark number sum
rules
\begin{align}
A_{1,0}^{I=0}(0) &= \int_0^1 dx\, \bigl( 
   u - \bar{u} + d - \bar{d} \,\bigr) = 3 \,, &
A_{1,0}^{I=1}(0) &= \int_0^1 dx\, \bigl( 
   u - \bar{u} - d + \bar{d} \,\bigr) = 1 \,,
\end{align}
in the proton, provided that $A_{1,0}^{I=0\slim (0)} = 3$,
$A_{1,0}^{I=1\slim (0)} = 1$ and $A_{1,0}^{I=0\slim (2,m)} =
A_{1,0}^{I=1\slim (2,m)} = 0$.

%%%%%%%%%%%%%%%%%%%%%%%%%%%%%%%%%%%%%%
\section{Summary}
\label{sect:sum}
%%%%%%%%%%%%%%%%%%%%%%%%%%%%%%%%%%%%%%

In this paper and its companion \cite{DMS2} we have calculated the
chiral corrections to the full set of twist-two generalized parton
distributions in the nucleon, using heavy-baryon chiral perturbation
theory.  For each form factor parameterizing the moments of these
distributions, our results include the order $O(q^2)$ relative to its
lowest-order expression.  We have presented a detailed account of the
power counting and of the operators that can contribute to the chiral
order we consider.  We find that the operator structure is relatively
simple in the basis of form factors specified by \eqref{ff-trafo} and
\eqref{ff-trafo-odd}.  With the exception of $\widetilde{M}_{n,k}$ and
$\widetilde{M}_{T n,k}$ only those pion-nucleon operator insertions
contribute to the loop corrections of a given form factor which
already provide its lowest-order expression at tree-level.
Furthermore, only operators with $\tau^A_{e+}$ or $\tau^A_{o+}$ from
\eqref{tau-even-def} and \eqref{tau-odd-def} contribute, but not those
with $\tau^A_{e-}$ or $\tau^A_{o-}$.  Our analysis also shows that
these simplifications will no longer hold at higher orders in the
chiral expansion.

Expressing our results in the basis of form factors parameterizing the
moments of the usual nucleon GPDs, we find that with the exception of
$A_{n,k}^{I=0}$ and $C_{n}^{I=0}$ all form factors receive chiral
corrections from loop graphs with nucleon operator insertions (see
Fig.~\ref{fig-1}).  They are of relative order $O(q^2)$ and contain
logarithmic terms $m^2 \log(m^2/\mu^2)$, but are independent of $t$
and of the indices $n,k$.  In several cases these corrections involve
a mixing between different form factors: $B_{n,k}$ receives
corrections involving not only its own lowest-order expression but
also the one of $A_{n,k}$, as seen in \eqref{res-gpd-vec}.  Likewise,
there are corrections to $\widetilde{B}_{n,k}$ from
$\widetilde{A}_{n,k}$, to $\widebar{B}_{T n,k}$ from $A_{T n,k}$, and
to $\widetilde{A}_{T n,k}$ from $A_{T n,k} + \frac{3}{2}
\widebar{B}_{T n,k}$.  We note that no such mixing occurs for the
linear combinations $A_{n,k} + B_{n,k}$ and $A_{T n,k} +
\widebar{B}_{T n,k}$.

Further corrections are due to loop graphs with pion operator
insertions (see Fig.~\ref{fig-2}a and b).  They only occur for form
factors which are accompanied by the maximal number of vectors
$\Delta_\mu$ in the decomposition of the associated matrix element, or
by one factor less.  Due to the quantum number restrictions for pion
operators, they only occur for even $n$ in the isosinglet and for odd
$n$ in the isotriplet sector.  Corrections starting at order $O(q)$
are obtained for $C_n^{I=0}$, $B_{n,n-1}^{I=1}$ and $\widetilde{A}_{T
n,n-1}^{I=1}$, and corrections starting at order $O(q^2)$ for
$B_{n,n-2}^{I=0}$, $\widetilde{A}_{T n,n-2}^{I=0}$, $\widetilde{B}_{T
n,n-1}^{I=0}$, $A_{n,n-1}^{I=1}$ and $\smash{\widebar{B}}_{T
  n,n-1}^{\slim I=1}$.
To order $O(q^2)$, the corrections for $C_n^{I=0}$ involve the
low-energy constants $c_1$, $c_2$, $c_3$ from the pion-nucleon
Lagrangian \eqref{LpiN}, whereas those for $B_{n,n-1}^{I=1}$ and
$\widetilde{A}_{T n,n-1}^{I=1}$ involve $c_4$.  The corrections from
pion operator insertions depend on $t$.  They are responsible for a
nonanalytic pion mass dependence of the derivatives of form factors at
$t=0$, namely a $1/m$ behavior for $\partial_t^{} C_n^{I=0}(0)$,
$\partial_t^{} B_{n,n-1}^{I=1}(0)$ and $\partial_t^{} \widetilde{A}_{T
n,n-1}^{I=1}(0)$ and a $\log(m^2/\mu^2)$ behavior in the other cases.
We note that these corrections also determine the onset of the
two-pion cut at timelike $t$ for the form factors in question.

The pseudoscalar form factors $\widetilde{B}_{n,n-1}^{I=1}$ receive
corrections from one-pion exchange (see Fig.~\ref{fig-2}c).  They take
the very simple form \eqref{Mnn-1-final} when expressed in terms of
physical quantities.  In particular, we find that the ratio
$\partial_t^{} \widetilde{B}_{n,n-1}^{I=1}(0) \slim\big/
\slim\partial_t^{} \widetilde{B}_{1,0}^{I=1}(0)$ of derivatives is
given by the moment $B_{n}^\pi$ of the pion distribution amplitude,
with corrections of order $m^3$.  It would be interesting to test this
prediction of chiral symmetry in lattice QCD calculations.

We have finally evaluated the corrections to the chiral-odd pion GPDs
at order $O(q^2)$, thus complementing the calculation \cite{DMS1} for
the chiral-even sector.  A compilation of our results for the values
and derivatives at $t=0$ of all moments of nucleon and pion GPDs is
given in Section~\ref{sect:gpd-results}.

%%%%%%%%%%%%%%%%%%%%%%%%%%%%%%%%%%%%%%
\section*{Acknowledgments}
%%%%%%%%%%%%%%%%%%%%%%%%%%%%%%%%%%%%%%

We are grateful to U.-G.\ Mei{\ss}ner for clarifying discussions.
This work is supported by the Helmholtz Association, contract number
VH-NG-004.

%%%%%%%%%%%%%%%%%%%%%%%%%%%%%%%%%%%%%

\end{document}